\documentclass[%
 reprint,
 amsmath,amssymb,
 aps,
]{revtex4-2}
\usepackage[dvipsnames]{xcolor}
\usepackage{braket} 
\usepackage{amsmath}
\usepackage{amsfonts}
\usepackage{amssymb}
\usepackage{hyperref}
\usepackage[singlelinecheck=false,justification=justified]{caption}
\usepackage{subcaption}
\usepackage{multirow}
\usepackage{graphicx}
\usepackage{dcolumn}
\usepackage{bm}
\usepackage[toc,page]{appendix}
\usepackage{titlesec}
\titlespacing\subsection{0pt}{12pt plus 4pt minus 2pt}{4pt plus 2pt minus 2pt}
\DeclareUnicodeCharacter{2212}{-}
\begin{document}

\preprint{APS/123-QED}

\title{Robust control and optimal Rydberg states for neutral atom two-qubit gates}
\author{Madhav Mohan$^1$}
\author{Robert de Keijzer$^1$}
\author{Servaas Kokkelmans$^1$}%
\affiliation{%
 $^1$Department of Applied Physics and Eindhoven Hendrik Casimir Institute, Eindhoven University of Technology, P. O. Box 513, 5600 MB Eindhoven, The Netherlands}%

\date{\today}
\begin{abstract}
We investigate the robustness of two-qubit gates to deviations of experimental controls, on a neutral atom platform utilizing Rydberg states. We construct robust CZ gates -- employing techniques from quantum optimal control -- that retain high Bell state fidelity $F > 0.999$ in the presence of significant deviations of the coupling strength to the Rydberg state. Such deviations can arise from laser intensity noise and atomic motion in an inhomogeneous coupling field. We also discuss methods to mitigate errors due to deviations of the laser detuning. The designed pulses operate on timescales that are short compared to the fundamental decay timescale set by spontaneous emission and blackbody radiation. We account for the finite lifetime of the Rydberg state in both the optimization and fidelity calculations -- this makes the gates conducive to noisy intermediate-scale quantum experiments, meaning that our protocols can reduce infidelity on near-term quantum computing devices.  
We calculate physical properties associated with infidelity for strontium-88 \mbox{atoms -- including} lifetimes, polarizabilities and blockade strengths -- and use these calculations to identify optimal Rydberg states for our protocols, which allows for further minimization of infidelity.
\end{abstract}
\maketitle

\section{\label{sec:level1}Introduction}
Neutral atoms trapped in optical tweezers are attractive candidates for applications in quantum information. These systems have proven scalable -- arrays of up to 324 atoms have been created \cite{PhysRevA.106.022611}, and large scale quantum simulations have been carried out \cite{ebadi_quantum_2021}. Quantum algorithms have also been demonstrated on neutral atom quantum computers \cite{graham_multi-qubit_2022}, and highly-entangled multiqubit states have been recently generated \cite{omranetal2019, graham_multi-qubit_2022}.\par
The problem of generating high-fidelity entanglement -- the fundamental resource towards constructing quantum circuits -- and quantum gates is a crucial one to the realization of full-fledged quantum computers. In particular, a two-qubit CZ gate combined with a full set of single-qubit operations is sufficient to perform universal quantum computation \cite{PhysRevA.52.3457}. 
Generating the necessary entanglement between neutral atoms involves excitation of the atoms using laser pulses, to electronic states of high principal quantum number $n$ -- termed \emph{Rydberg} states. Using such a scheme, high-fidelity entangling operations between two qubits (with Bell state fidelity $>0.991$) have been recently demonstrated by  Madjarov \emph{et al.} \cite{madjarov_high-fidelity_2020}. Theoretical predictions of two-qubit gate fidelities $F > 0.9999$ \cite{shannon2021, PhysRevA.94.032306} show that high-fidelity gates are possible on this platform.

A high fidelity of logic gates is of importance as the fidelity limits the depth of quantum circuits that can be constructed, and thus constrains the complexity of the algorithms that can be implemented. Furthermore, strict requirements are placed on gate fidelities and qubit coherence times by quantum error correction \cite{Devitt_2013}, and current experimental demonstrations -- while highly promising -- fall short of these requirements.
This era of quantum computing is termed the noisy intermediate scale quantum (NISQ) era \cite{Preskill2018quantumcomputingin}, where multiple sources of decoherence and infidelity are present, that must be carefully accounted for and mitigated. Following the seminal proposal for realizing CZ gates with Rydberg atoms \cite{PhysRevLett.85.2208}, various two-qubit gate protocols have been proposed, that promise high fidelities. Gates that are robust to deviations of controls have been proposed, using analytical schemes based on adiabaticity \cite{mitra_robust_2020, saffman_quantum_2016}. Time-optimal gates -- gates that minimize the duration of operation, thus mitigating error sources that propagate in time -- have recently been identified \cite{jandura_time-optimal_2022, pagano_error_2022}. These protocols were devised using quantum optimal control (QOC) -- a class of methods that aim to find optimal ways to steer a quantum system from an initial state to a desired final state using time-dependent controls \cite{Koch_2016}. Using QOC methods, quantum gates have been engineered and experimentally demonstrated on superconducting platforms \cite{heeres_implementing_2017, 9835639, werninghaus_leakage_2021, PhysRevLett.125.170502}, and non-classical states have been generated on the Rydberg platform \cite{PhysRevX.10.021058, omranetal2019}.  \par
Although such Rydberg experiments often employ alkali atoms, recent quantum information architectures have been identified for alkaline-earth atoms \cite{Weggemans2022solvingcorrelation, PhysRevLett.101.170504, PhysRevA.105.052438}. Notably, the high-fidelity two qubit entanglement study of Madjarov \emph{et al.} employs the metastable clock state of strontium-88 ($^{88}$Sr) atoms along with the electronic ground state, resulting in a long-lived qubit configuration \cite{madjarov_entangling_2021}. Moreover, the rich electronic structure of these atoms allows for novel applications, and is expected to pave the path towards higher gate fidelities -- and in the longer term, fault-tolerant quantum computing \cite{PhysRevX.12.021049}. \par
In this work, we improve upon the current state of the art in two ways, relating to two-qubit gate design and the choice of Rydberg state.\par
First, we use QOC methods to investigate the design of two-qubit gates operating on a timescale comparable to the time-optimal gates, that are also robust to deviations of \emph{Rabi frequency}, which quantifies the coupling strength from the computational basis states to the Rydberg state. These deviations arise due to laser intensity noise \cite{madjarov_high-fidelity_2020} and motion of the atom in a spatially inhomogeneous coupling laser field \cite{mitra_robust_2020}. 
We achieve excellent robustness for gates that are near time-optimal and thus also mitigate errors due to deviations of the detuning, such as those arising from thermal Doppler shifts \cite{mitra_robust_2020} and stray electric fields in the experiment. The corresponding pulses are smooth, and adhere to realistic experimental constraints on Rabi frequency and detuning, achievable in current experimental setups. \par
The second way in which we improve upon the state of the art, is with the identification of optimal Rydberg states in $^{88}$Sr atoms, in order to minimize the infidelity of our gate protocol. We achieve this by accounting for various sources of error -- including spontaneous emission from the Rydberg state, blackbody radiation (BBR) and sensitivity to stray electric fields in the experiment -- and investigating the scaling of these errors with $n$.\par
The methods we develop for these results can be adapted to other Rydberg atom experiments, and thus provide a closer link between theoretical advances and experimental performance of quantum computers.\par  
This work is structured as follows. 
In Sec.~\ref{theory} we motivate the computation of atomic properties and formulate both the two-qubit gate dynamics, and the variational framework.  
In Sec.~\ref{robustsection}, we discuss the results of our optimization, present two new CZ gate protocols and compare them to the time-optimal gates. Section ~\ref{whatstate} identifies the regime of optimal Rydberg states to minimize the infidelity of our protocols.
\section{Theory and problem formulation}
\label{theory}
\begin{figure}
\includegraphics[width=5cm]{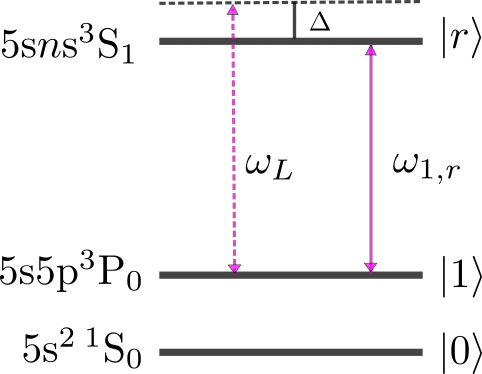}
\caption{Energy level diagram of the $^{88}$Sr states relevant for our protocol. The $\ket{1}$ state is coupled to the $\ket{r}$ state with the laser field. The detuning of the laser $\Delta$ is the difference of the frequency of this transition $\omega_{1,r}$ with the laser frequency $\omega_L$. }
\label{fig1}
\end{figure}
\subsection{Modeling of atomic properties}
\label{atoms}
In Rydberg platforms, one encodes quantum information in the electronic or nuclear states of an atom. The discussion presented here considers strontium-88 atoms, but protocols for atoms such as rubidium-87 \cite{PhysRevLett.121.123603}, ytterbium-171 \cite{PhysRevA.105.052438} and cesium-133 \cite{mitra_robust_2020} have been studied. \par Our protocol employs the electronic ground state \mbox{$5\mathrm{s}^2$  $^1\mathrm{S}_0$} and the long-lived metastable clock state \mbox{$5\mathrm{s}5\mathrm{p}\,^3\mathrm{P}_0$} as qubit states $\ket{0}$ and $\ket{1}$ respectively. Entanglement is facilitated with the Rydberg state 
 $\ket{r}$, $5\mathrm{s}n\mathrm{s}\,^3\mathrm{S}_1$. Figure ~\ref{fig1} presents the main transitions of interest in our system. Single-photon transitions are achievable from the $\ket{1}$ state to the $\ket{r}$ state \cite{madjarov_high-fidelity_2020}, which eliminates decay from intermediate states -- a source of infidelity common to many Rydberg experiments \cite{saffman_quantum_2016}.  \par

The physical properties of atoms relevant to this discussion can be captured with dipole matrix elements. The dipole matrix element $d_{a,b}$ between two states $a$ and $b$ is $\bra{b}e\mathbf{x}\ket{a}$, where $e$ is the electronic charge and $\mathbf{x}$ the position vector of the active electron. Computing these elements is crucial to our work, as it allows us to calculate lifetimes of the Rydberg states, interaction strengths and polarizabilities, and their scaling with the principal quantum number of the Rydberg state, $n$. How these quantities translate to infidelities is discussed in Sec. \ref{whatstate}. \par One can capture the physics of interest for Rydberg states with modified eigenstates of Hydrogen atoms. The energy levels for a series of Rydberg states is given as \cite{Seaton_1983}
\begin{equation}
E_n = -\frac{1}{2(n - \delta_{lj}(n))^2} = -\frac{1}{2(n^*)^2},
\end{equation}
where the quantum defect $\delta_{lj}$ captures the deviation from the hydrogenic states, $n^*$ is the \emph{effective principal quantum number}, $l,j$ are the orbital angular momentum and the total angular momentum numbers of the valence electron, and atomic units are used. 
\par The scaling of the relevant physical properties with $n$ is well understood for alkali atoms.
For instance, the lifetime due to spontaneous emission, $\tau_{sp}$ of Rydberg states in alkali atoms scales as $(n^*)^3$ \cite{gounand1979}, whereas the BBR lifetime $\tau_{BBR}$, as well as the total lifetime $(1/\tau_{sp} + 1/\tau_{BBR})^{-1}$ (see Eq.~\ref{eqn:totallifetime}) both scale as $(n^*)^2$\cite{PhysRevA.79.052504, ediss18152}.  \par For alkaline-earth atoms, the presence of two valence electrons can complicate the physics \cite{Vaillant_2014}, although most of the physical quantities relevant to our study are straightforward to compute. The corresponding calculations are carried out and detailed in Appendix \ref{section:appendixB}.\par 
\subsection{Dynamics and two-qubit gates}
\label{twoqubitgates}
For a single qubit -- with the $\ket{1}$ state coupled to the Rydberg state $\ket{r}$ by a laser field -- the Hamiltonian is given by
\begin{equation}
\label{eqn:1q}
H_{1q} = \frac{\Omega(t)}{2}(\ket{1}\bra{r} + \ket{r}\bra{1}) - \Delta(t)\ket{r}\bra{r},
\end{equation} 
where $\Omega$ is the Rabi frequency and $\Delta = \omega_{1,r} - \omega_L$ the detuning of the laser respectively (with $\omega_{1,r}$ the frequency of the transition and $\omega_L$ the laser frequency)--- see Fig.~\ref{fig1}. The Rabi frequency $\Omega \propto \sqrt{P} \bra{r}ex\ket{1}$, where $P$ is the total power of the driving laser and $\bra{r}ex\ket{1}$ is the radial dipole matrix element (RDME) (see Eq.~\ref{eqn:rdme}) between the clock state and the Rydberg state. In this context, a pulse is the profile of $\Omega(t)$ and $\Delta(t)$ in time.
\par The Hamiltonian for a two-qubit system with both atoms coupled to the same laser field is
\begin{equation}
\label{eqn:2q}
H_{2q} = H_{1q}\otimes I + I \otimes H_{1q} + V_{\mathrm{int}}\ket{rr}\bra{rr},
\end{equation} 
where $V_{\mathrm{int}}$ is the interaction strength between two atoms excited to Rydberg state $\ket{r}$. In our protocol we utilize the van der Waals interaction, $V_{\mathrm{int}} = -C_6/R^6$, where $R$ is the interatomic distance and $C_6$ a state-dependent coefficient.\par   
The dynamics of the Hamiltonian in Eq.~\ref{eqn:2q} can be employed to generate two-qubit quantum gates. We work in the regime of the \emph{Rydberg blockade} -- where the interaction strength between the two atoms is much larger than the Rabi frequency, such that an atom excited to state $\ket{r}$ suppresses the excitation of adjacent atoms to $\ket{r}$ \cite{PhysRevLett.85.2208}. While several protocols assume $V_{\mathrm{int}} = \infty$, this can lead to errors due to the finite blockade strength of Rydberg states -- hence, we include the finite blockade strengths in our optimization. For the Rydberg states considered in Sec. \ref{whatstate}, we observe no infidelity arising from the finite strengths, and consistently find gates with $(1 - F) \sim 10^{-5}$.\par 
The general class of protocols enabled by Eq.~\ref{eqn:2q} in the blockade regime has been well studied, and is captured by the dynamics of the computational states. The $\ket{0}$ state -- and consequently, the $\ket{00}$ state -- is dark to the laser field and is not affected in our protocol. The dynamics of the $\ket{01}$ and $\ket{10}$ states can thus be evaluated by considering the atoms individually, and reducing them to two-level systems $\ket{1}, \ket{r}$ with Rabi frequency $\Omega(t)$. Further, the dynamics of $\ket{01}$ and $\ket{10}$ are identical if the pulse is \emph{symmetric} i.e., the same laser field is applied to both the atoms, which is the case in our protocol. In the limit of infinite blockade, the dynamics of the $\ket{11}$ state can also be described by a two-level system of $\ket{11}, \ket{b}$ for the bright state $\ket{b} = (\ket{1r} + \ket{r1})/\sqrt{2}$ with Rabi frequency $\sqrt{2}\Omega(t)$ \cite{mitra_robust_2020}. This description is valid as the large $V_{\mathrm{int}}$ suppresses excitations to the $\ket{rr}$ state, and is valuable to understand how entanglement is generated. However, in order to account for finite $V_{\mathrm{int}}$, we will consider the three-level system $\ket{11}, \ket{b}, \ket{rr}$.\par The dynamics mentioned above lead to accumulation of phases on the states -- these are state-dependent, due to the different Rabi frequencies. By setting a constraint on the final phases, and applying an additional single-qubit rotation, one can realize CPHASE gates \cite{pagano_error_2022} characterized by the transformation $\ket{00} \to \ket{00}, \ket{01} \to \ket{01}, \ket{10} \to \ket{10}, \ket{11} \to \mathrm{e}^{i\phi}\ket{11}$. At $\phi = \pi$, the CZ gate is obtained.\par  To account for various sources of error, corrections can be made to Eqns. \ref{eqn:1q} and \ref{eqn:2q}. The finite lifetime of the Rydberg state is accounted for by including the term $H_{\mathrm{decay}} = -i\frac{\Gamma}{2}\ket{r}\bra{r}$ in Eq.~\ref{eqn:1q}, with the total decay width $\Gamma$ defined in Eq.~\ref{eqn:totallifetime}. This assumes that the decay occurs only to states outside of the computational subspace, and thus slightly overestimates the error \cite{PhysRevA.96.042306}. For the $5\mathrm{s}n\mathrm{s}^3\mathrm{S}_1$ Rydberg state in $^{88}$Sr, branching ratios of the decay processes have been calculated, that verify the validity of this assumption \cite{madjarov_entangling_2021}.   
Deviations of the Rabi frequency and detuning, $\delta\Omega, \delta\Delta$ are included by making the substitutions $\Omega(t) \to \Omega(t) + \delta\Omega$ and $\Delta(t) \to \Delta(t) + \delta\Delta$ in Eq.~\ref{eqn:1q}. We thus focus on shot-to-shot deviations, and discuss the potential time-dependent deviations in Appendix \ref{appendix:offresonant}. \par Another question of interest for experiments is the choice of Rydberg state that minimizes the infidelity of a gate. This question has been considered in some detail. For higher-lying states, we expect higher infidelity arising from the sensitivity of $\ket{r}$ to stray electric fields \cite{saffman_quantum_2016}, but a decrease in the infidelity arising from decay of $\ket{r}$. The scaling of intrinsic gate error with $n$, for the gate proposed by Jaksch \emph{et al.} \cite{PhysRevLett.85.2208} was presented for Rubidium \cite{RevModPhys.82.2313}. However, how the physical quantities translate into errors is protocol dependent and it is difficult to interpret the results in context of the robust gates we will obtain, or of other recent gates proposed. \par Further, as we consistently find high-fidelity pulses in the presence of finite blockade strengths for the $^{88}$Sr Rydberg states considered, the discussion by Saffman \emph{et al.} \cite{RevModPhys.82.2313} of mitigating this error by going to a higher $n$ -- or varying interatomic distances -- does not apply here. We also safely ignore the effects of off-resonant scattering in our $\mathrm{5s5p^3P_0}\leftrightarrow\mathrm{5s}n\mathrm{s^3S_1}$ transition, since these rates are negligible with respect to the pulse durations, as detailed in Appendix~\ref{appendix:offresonant}.

\subsection{Variational optimization}
\label{QOC}

We now discuss the dynamics formulation of Sec. \ref{twoqubitgates} in the context of quantum optimal control (QOC), towards variational optimization. \par As mentioned in Sec.\ref{sec:level1}, time-optimal CZ and $\mathrm{C_2Z}$ gates -- aiming to find the fastest possible gate protocol -- have been obtained using QOC for a Rydberg atom platform \cite{jandura_time-optimal_2022}.$\;$In the rest of this work, we will compare and contrast our robust gates to these specific time-optimal gates.$\;$ Analytic protocols which are robust to deviations of experimental controls have also been proposed \cite{saffman_symmetric_2020, mitra_robust_2020} -- it is thus a worthwhile question whether robustness can be included in a variational optimization framework. \par
Numerical robustness has been studied for two-qubit gates on the Rydberg platform before \cite{PhysRevA.90.032329, PhysRevA.102.042607} -- and while these studies demonstrate excellent robustness is possible, we note certain differences. For instance, robust gates were obtained with single-site addressability, for atoms with two-photon transitions to $\ket{r}$ \cite{PhysRevA.90.032329} -- the control parameters of the obtained pulses oscillate on a time scale that is difficult to implement experimentally. In comparison to these pulses, the pulses we find are smooth, and do not require single-site addressability, significantly easing experimental implementation \cite{saffman_symmetric_2020}. As these are optimized for one-photon transitions in $^{88}$Sr, the fidelities $F$ are inherently higher, but also cannot be directly compared to fidelities of the two-photon gate as the experimental setups considered differ.\par 

Towards the optimization, we employ the augmented Lagrangian trajectory optimizer (ALTRO) method \cite{Howell-2019-122091}, used before to study robust QOC for superconducting qubits \cite{propson_robust_2022}. Notably, ALTRO can satisfy constraints up to a specified tolerance by using an augmented Lagrangian method (ALM) to adaptively adjust Lagrange multiplier estimates for the constraint functions. This makes the optimizer well suited for our problem, which considers multiple constraints. In this method, one formulates the problem to be solved as a trajectory optimization problem \cite{propson_robust_2022}. The cost function we use is 
\begin{equation}
\label{costfn}
J(\mathbf{c}) = (\mathbf{x}_N - \mathbf{x}_T)^{\intercal} Q_N (\mathbf{x}_N - \mathbf{x}_T) + \sum_{k = 1}^{N - 1}\mathbf{c}_k^{\intercal} R_k \mathbf{c}_k,
\end{equation}
where the augmented control vector $\mathbf{c}$  encapsulates the experimental controls of our system, $\mathbf{x}$ is the augmented state vector consisting of the variables the control can influence -- such as the state populations and phases, $\mathbf{x}_T$ denotes the target augmented state, $\mathbf{x}_N$ is the augmented state vector at the final time step and the sum runs over time steps of the discretized problem. $R_k$ and $Q_N$ are diagonal weight matrices on the control at each time step $k$, and on the augmented state at the final time step $N$, respectively. Such a cost function can be efficiently computed with ALTRO \cite{Howell-2019-122091, propson_robust_2022}. The number of time steps $N = 100$ for all the results presented in this paper, and we observe negligible errors due to this discretization. \par
With one notable exception discussed later on, the augmented controls and states used in this work are 
\begin{equation}
\label{formulation}
\mathbf{c} = 
\begin{pmatrix}
\dot{\Omega} \\[7pt] \dot{\Delta}
\end{pmatrix},\;\;\;
\hat{\mathbf{x}} = 
\begin{pmatrix}
\partial_{\Omega}\Psi\\[6.5pt]
\partial_{\Delta}\Psi\\[6.5pt]
\phi\\[6.5pt]
P_{\mathrm{tot}}\\[6.5pt]
\overline{T}_{\ket{r}}
\end{pmatrix},
\end{equation}
where $\Psi = (\Psi^{01}, \Psi^{11})$  are the time-dependent wavefunctions associated with initial computational basis states $\ket{01}$ and $\ket{11}$, $\phi$ is the CPHASE gate angle as introduced in Sec. \ref{twoqubitgates}, $P_{\mathrm{tot}}$ refers to the total population in the computational basis states and $\overline{T}_{\ket{r}}$ is the average integrated Rydberg lifetime (see Eq.~\ref{avgtime}). We use the notation $\dot{a} = da/dt$, and $\partial_y a = \partial a/\partial y$ for some parameter $y$. $\hat{\mathbf{x}}$ is the reduced state vector and consists of the terms in $\mathbf{x}$ that are penalized in Eq.~\ref{costfn} -- for the remaining terms, the corresponding element in $Q_N$ is set to zero. The complete augmented state vector $\mathbf{x}$ is given and motivated in Appendix \ref{qocappendix}. In Eq.~\ref{formulation}, it is implicit that $\mathbf{c}$ has different elements $\mathbf{c}_k$ at each time step $k$. The dynamics of the system are governed by the time-dependent Schr\"{o}dinger equation.\par
Below, we motivate our formulation of the problem and the individual terms in Eq.~\ref{formulation}. As smooth pulses are more feasible from an experimental point of view \cite{saffman_symmetric_2020}, we set $\dot{\Omega}$ and $\dot{\Delta}$ as the augmented controls -- instead of $\Omega$ and $\Delta$ -- in order to penalize pulses with discontinuous jumps, with the weights $R_k$ in Eq.~\ref{costfn}.\par For a full description of the dynamics, we would have to consider the wavefunctions associated with all basis states, $\Psi^{00}, \Psi^{01}, \Psi^{10}$ and $\Psi^{11}$. We can, however, leave out $\Psi^{00}$ as this state is not coupled to the Rydberg states in our protocol. Furthermore, as the protocol is symmetric, $\Psi^{01}$ and $\Psi^{10}$ adhere to the same dynamics -- thus we can reduce the computational expense by optimizing for only one of the two basis states in this formalism.\par There are various ways to encode robustness into the optimization objective \cite{PhysRevA.90.032329, PhysRevA.102.042607}. We do so by including $\partial_{\Omega}\Psi$ and $\partial_{\Delta}\Psi$ as augmented states, in order to penalize the sensitivity of the desired wavefunction to deviations of the experimental parameters $\Omega$ and $\Delta$ respectively. This method -- termed the derivative method -- is intuitive and computationally inexpensive. For single-qubit gates on the superconducting platform, it achieves significantly higher robustness at lower computational cost compared to other available methods \cite{propson_robust_2022}. \par$\phi_{\ket{11}}$, and $\phi_{\ket{10}} = \phi_{\ket{01}}$ denote the phases accumulated by the computational states at the end of the gate. Enforcing the condition 
\begin{equation}
\label{phaseeqn}
\phi = \phi_{\ket{11}} - 2\phi_{\ket{01}} = (2m + 1)\pi,
\end{equation}
for $m$ an integer, results in a CZ gate \cite{pagano_error_2022}, up to one-qubit rotations. High fidelity in the noiseless case is ensured by including $\phi$ in $\mathbf{x}$ and $\pi$ as the corresponding element in $\mathbf{x}_T$ -- such that the condition in Eq.~\ref{phaseeqn} is ensured -- and by including $P_{\mathrm{tot}}$ (see Eq.~\ref{populationtotal}) in $\mathbf{x}$ such that the controls return the computational basis states to themselves (up to the acquired phase), in the final augmented state $x_{N}$. \par $\overline{T}_{\ket{r}}$ captures the total time spent in the Rydberg state, averaged across the computational basis states -- and Sec. \ref{robustsection} uses this as a metric to compare decay errors from $\ket{r}$ between different protocols. For a given configuration, minimizing $\overline{T}_{\ket{r}}$  reduces the error due to spontaneous emission and BBR. \par To ensure that the solver does not trade optimization of $\phi$ and $P_{\mathrm{tot}}$ with other terms in $\mathbf{x}$, we also include the appropriate constraints for these terms (treated with the ALM method) in the variational optimization. Constraints on Rabi frequency and detuning are also implemented, and will be discussed in Sec. \ref{robustsection}.\par 
To the best of our knowledge, our study is the first to use the derivative method \cite{propson_robust_2022} to study robustness both on the Rydberg platform and for two-qubit gates -- we expect use case scenarios for robust state preparation, multiqubit gates and quantum algorithms inspired by QOC \cite{rvdk}.
\section{Constraint-specific robust gates}
\label{robustsection}
\begin{figure}
     \centering
     \begin{subfigure}[b]{0.41\textwidth}
         \centering
         \includegraphics[width=\textwidth]{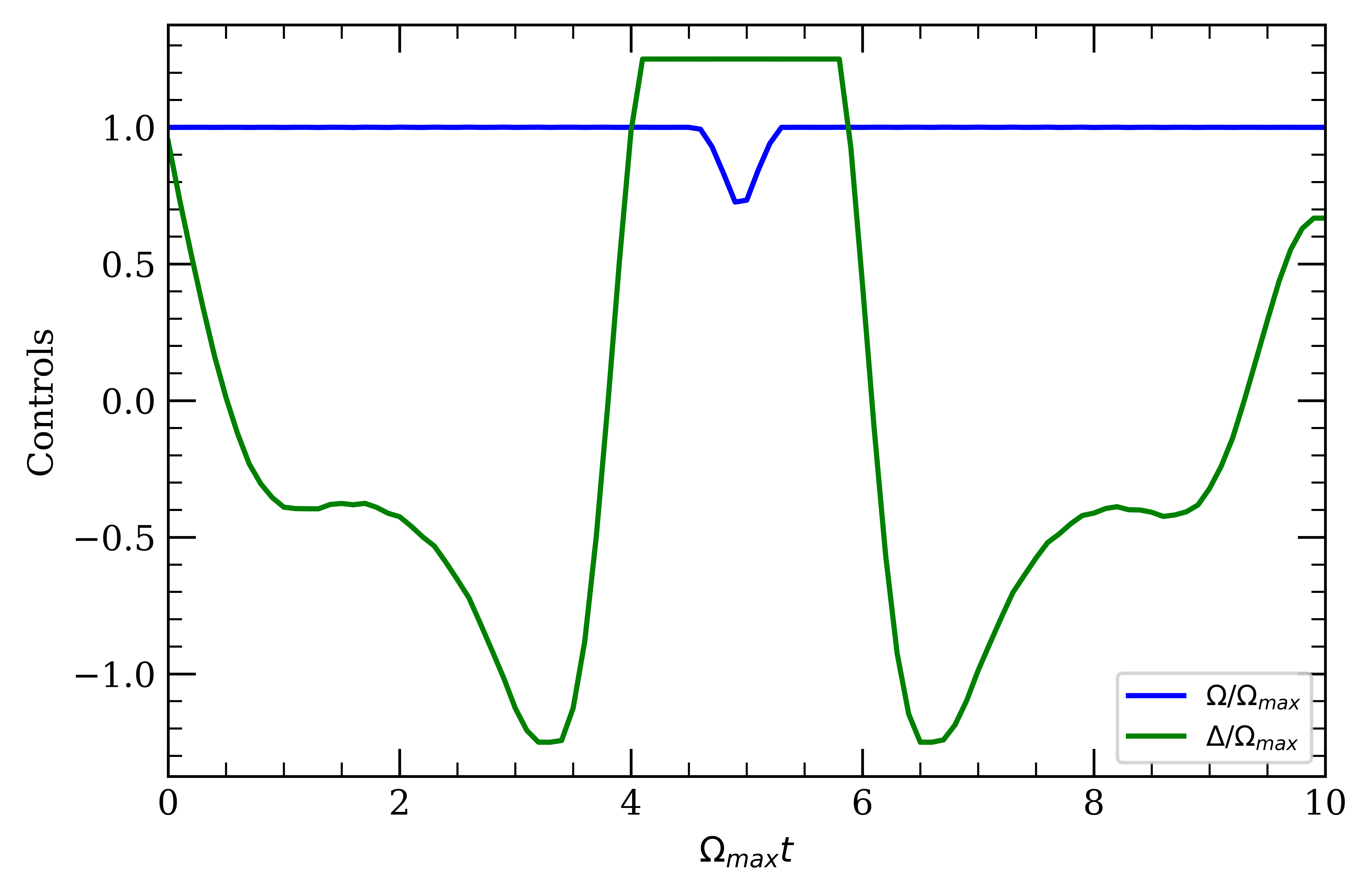}
         \caption{Protocol A : $T = 10/\Omega_{\mathrm{max}}, |\Delta(t)|\leq1.25\Omega_{\mathrm{max}}$}
         \label{pulseplot_protocolA}
     \end{subfigure}
     \hfill
     \begin{subfigure}[b]{0.41\textwidth}
         \centering
         \includegraphics[width=\textwidth]{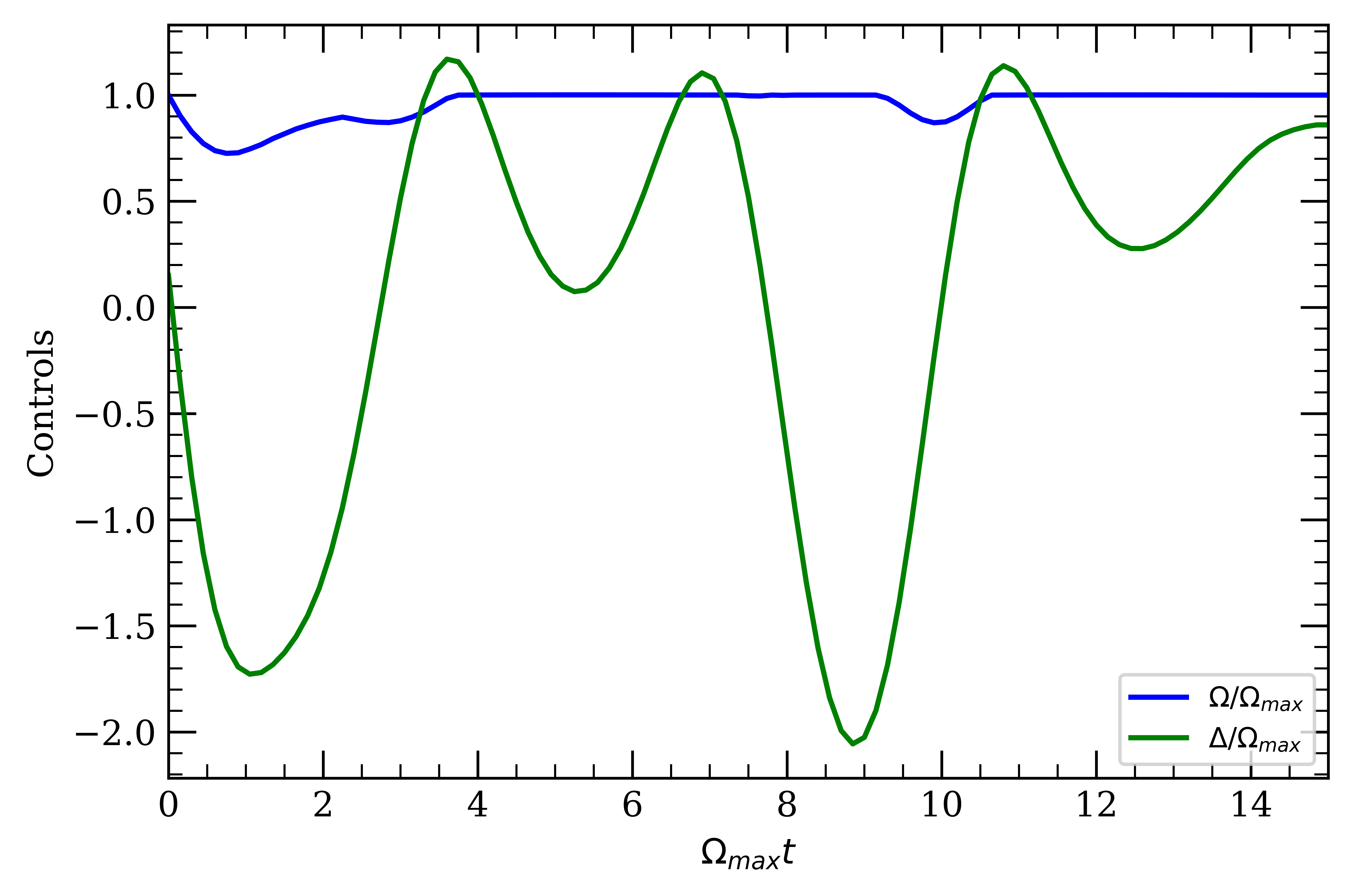}
         \caption{Protocol B : $T = 15/\Omega_{\mathrm{max}}, |\Delta(t)|\leq2.25\Omega_{\mathrm{max}}$}
         \label{pulseplot_protocolB}
     \end{subfigure}
     \hfill
       \caption{Two protocols to implement robust CZ gates. We highlight the smoothness of the pulses, the fast time scales (compared to decay from $\ket{r}$, see Eq. \ref{bigeqn}), and realistic constraints on detuning, ensuring that the gates can be implemented on current experimental setups.}
        \label{pulseplots}
\end{figure}

We now discuss the results of the optimization as detailed in Sec. \ref{QOC}. We will quantify the performance of our gates by computing the Bell state fidelities $F$. As discussed, in the absence of sources of error -- except finite $V_{\mathrm{int}}$, which is always included in our optimization -- tight constraints are implemented to ensure infidelity of the Bell state $(1 - F) \sim 10^{-5}$.  
While the results of this section can be understood by expressing quantities in units of the maximum Rabi frequency $\Omega_{\mathrm{max}}$, we set $\Omega_{\mathrm{max}} = 2\pi \times 6.8$ MHz -- the Rabi frequency demonstrated for the $n = 61$ state in $^{88}$Sr \cite{madjarov_high-fidelity_2020} -- to provide a better understanding of the physical values and time scales. We set the lifetime of $\ket{r}$ as $\tau_{n=61} = 96.5\mathrm{\mu s}$ -- the calculation of the lifetime is detailed in Appendix \ref{section:appendixB}. Unless specified otherwise, the fidelities in this section account for decay from the Rydberg state and finite blockade strengths. \par We use the ARC3.0 library \cite{ROBERTSON2021107814} to compute the van der Waals coefficient $C_6 = -2\pi \times 181$GHz $\mu\mathrm{m}^6$ -- we find the $n = 61$ state is well-resolved above $ R \approx 2\mu$m. The interatomic axis is taken perpendicular to the quantization axis of $\ket{r}$. We set $R = 3.5\mu$m in the simulation, both for the optimizations and fidelity calculations -- an experimentally feasible value -- and find negligible errors from the finite blockade strength. The physical parameters used in our simulation are collected in Table \ref{table:1}. 
To provide context, we will also compare our results to the time-optimal pulses \cite{jandura_time-optimal_2022, pagano_error_2022}. 
\par To illuminate our method, the role of constraints and relevant parameters (as introduced in Sec. \ref{QOC}), we present two gates robust to deviations of Rabi frequency $\delta\Omega$ -- Protocols A and B -- in Fig.~\ref{pulseplots}.\par
\begin{figure}
\includegraphics[width=0.475\textwidth]{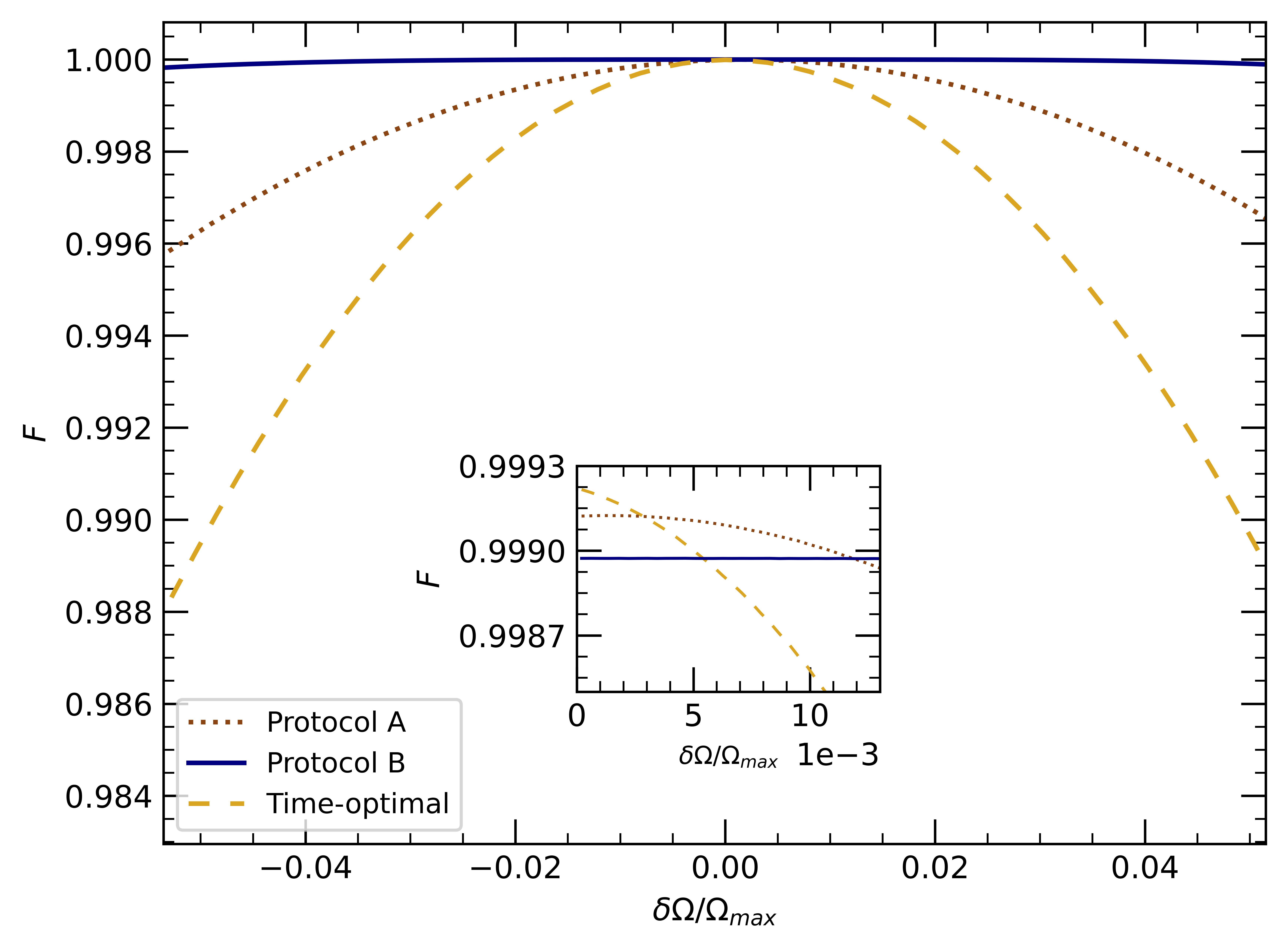}

\caption{Fidelities $F$ of the protocols from Fig.~\ref{pulseplots} against variations in $\delta\Omega$, plotted in the absence of other error sources. Compared to the time-optimal pulses, we observe significantly enhanced robustness for both protocols that we propose. Protocol A retains $F \sim 0.996$ and B retains $F > 0.9997$ for the range of variations plotted here. Inset -- we now consider decay from $\ket{r}$ and look at small Rabi variations -- for $\delta\Omega/\Omega_{max} \sim 0.5\%$ both protocols achieve higher fidelity than the time-optimal pulses.}
\label{robust12}
\end{figure}
The protocols are differentiated by the implemented constraints, gate times and states included in Eqns. \ref{costfn} and \ref{formulation}. 
Protocol A implements the constraint $|\Delta(t)|\leq1.25\Omega_{\mathrm{max}}$, has a total pulse duration of \mbox{$T_{A} = 10/\Omega_{\mathrm{max}} (\approx 0.23 \mathrm{\mu s})$} and accounts for robustness to deviations of detuning $\delta\Delta$ (see Sec. \ref{QOC}). This time scale is comparable to the time-optimal protocols ($T_{\mathrm{opt}} = 7.61/\Omega_{\mathrm{max}}$) \cite{jandura_time-optimal_2022} -- and thus benefits from the associated mitigation of errors that propagate in time. This protocol is intended to be implementable for current experiments. It places modest requirements on the experimental controls and is near time-optimal.\par For protocol B, we relax the constraints to $|\Delta(t)|\leq 2.25\Omega_{\mathrm{max}}$, set a pulse duration of $T_B = 15/\Omega_{\mathrm{max}}$ and do not include $d\Psi/d\Delta$ in the optimization. Protocol B is intended to probe the limits of our optimization -- to explore how robust CZ gates can be made to $\delta\Omega$ deviations, while still operating on a short timescale. During our optimization, we noticed that the optimizer tends to reach high values of $\Delta$. Hence, even for protocol B we find it useful to set bounds on $|\Delta(t)|$ -- significant robustness is already achieved for protocol B with bounds set to $|\Delta| < 2.25\Omega_{\mathrm{max}}$, and we notice negligible improvements over this protocol on relaxing the constraints further. \par From Fig.~\ref{robust12}, we note significant improvements in robustness of both protocols A and B compared to the time-optimal pulses.\par At $\delta\Omega = \pm0.05\Omega_{\mathrm{max}}$, the time-optimal pulse fidelity drops to $F_{\mathrm{opt}} = 0.9896$, while the fidelity of protocol A, $F_A = 0.9964$, and the fidelity of protocol B, $F_B = 0.9989$ -- these values also account for decay from $\ket{r}$. Hence, both protocols allow us to achieve high-fidelity gates in the presence of significant $\delta\Omega$ deviations and decay.\par
\begin{figure}
 	\centering
     \begin{subfigure}{0.495\textwidth}
     \caption{}
         \centering
         \includegraphics[width=\textwidth]{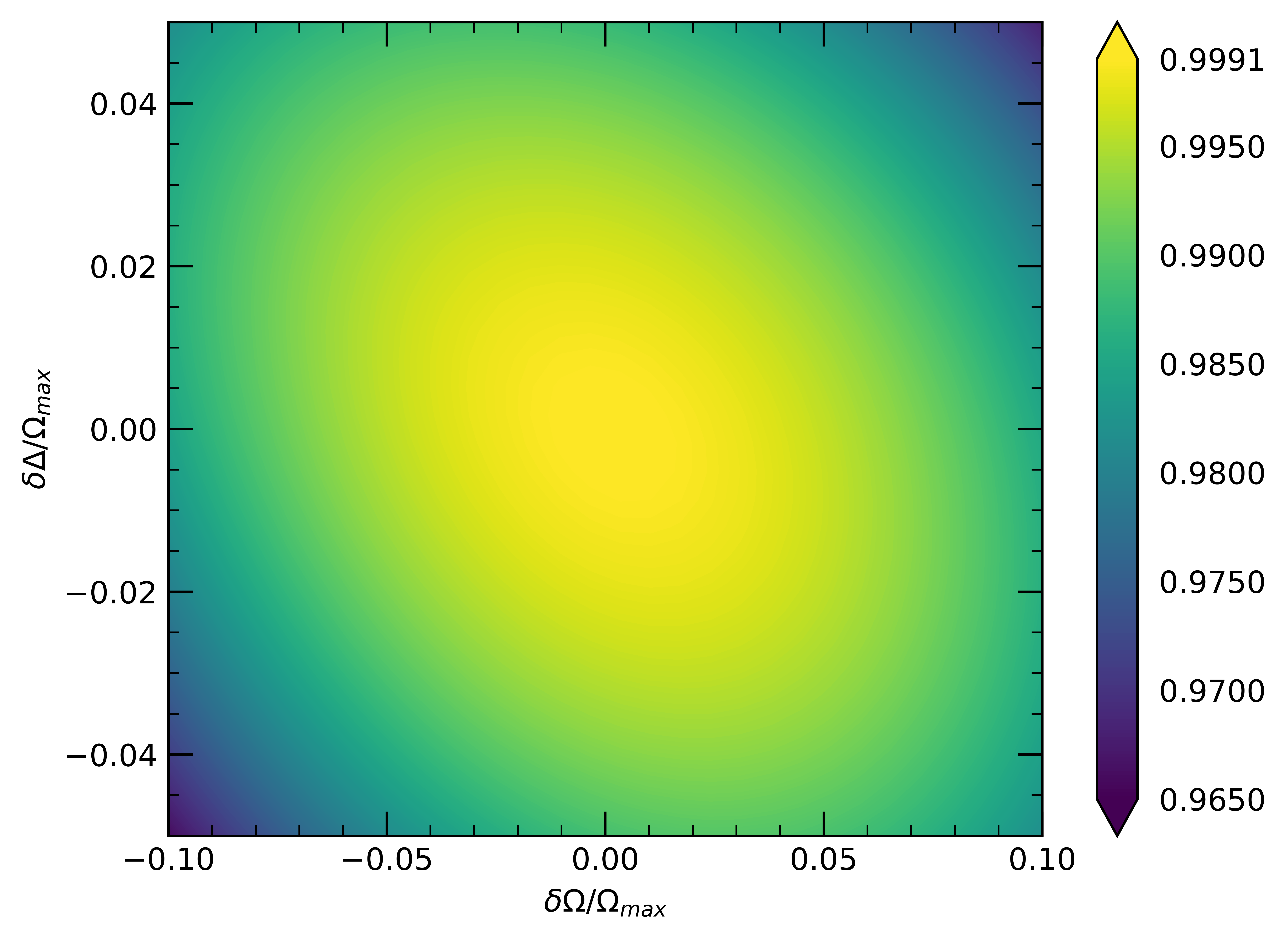}
         \label{robustplot_protocolA}
     \end{subfigure}
     \hfill
     \begin{subfigure}{0.495\textwidth}
         \centering
         \caption{}
         \includegraphics[width=\textwidth]{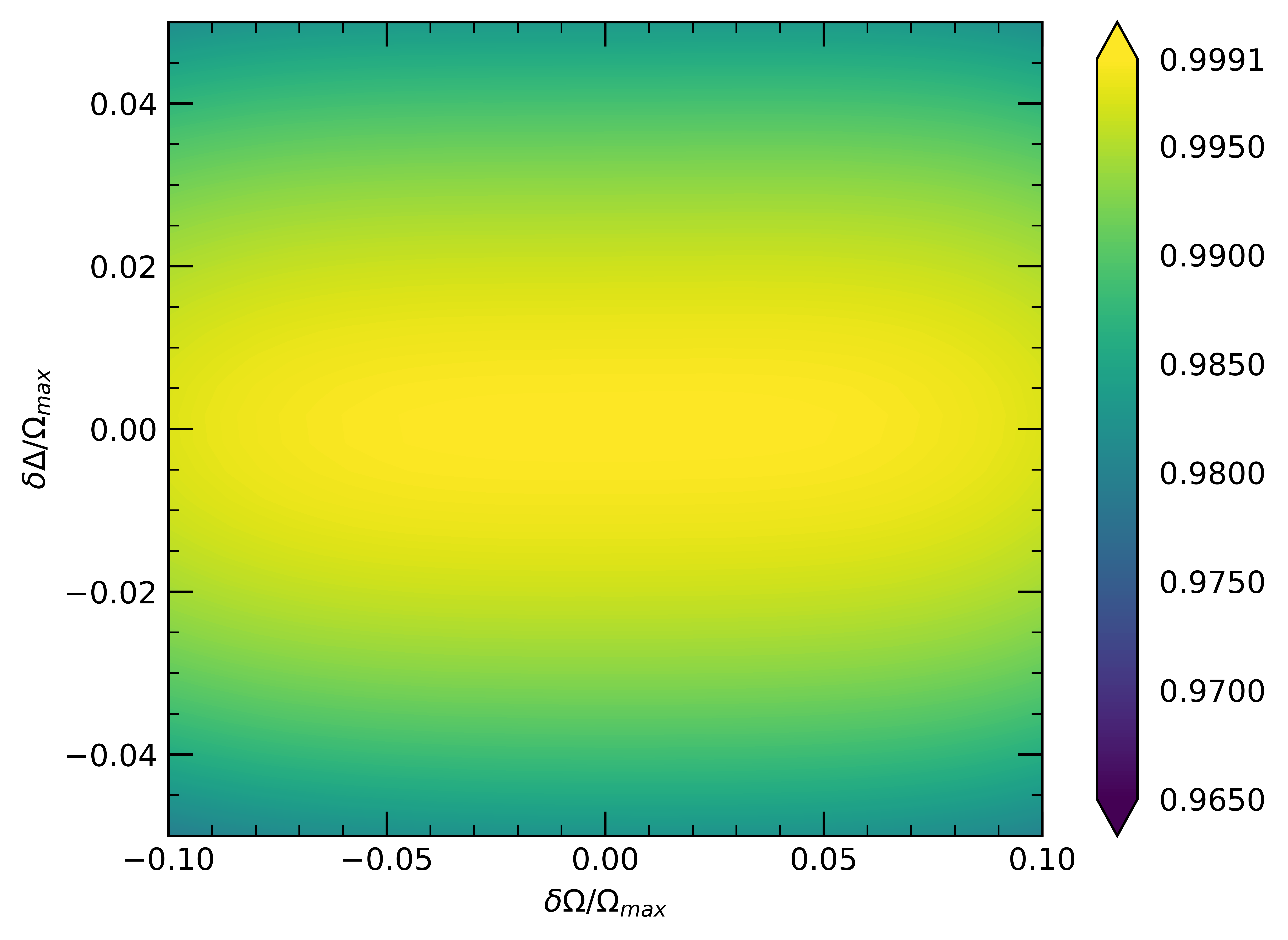}
         \label{robustplot_protocolB}
     \end{subfigure}
     \hfill
        \caption{Contour plot of fidelity $F$ to investigate the robustness of the proposed pulses against $\delta\Omega, \delta\Delta$. \\(a) Protocol A is slightly more robust to $\delta\Delta$ than (b) protocol B, whereas B is significantly more robust to $\delta\Omega$. Both pulses lead to high-fidelity ($F \sim 0.999$) gates across a wide range of deviations.}
        \label{robustplots}
\end{figure}

We find that pulses A and B still offer a significant advantage for minor deviations of $\Omega$ -- at $\delta\Omega = 5\times10^{-3}\Omega_{\mathrm{max}}$ ($0.5\%$ of the maximum Rabi frequency) both protocols A and B have higher fidelity $F\sim0.999$ than the time-optimal pulse (see inset of Fig.~\ref{robust12}). The robustness of the protocols -- in the presence of decay from $\ket{r}$ -- to $\delta\Omega$ and $\delta\Delta$ deviations is studied in Fig.~\ref{robustplots}.\par 
For the time-optimal gates, the average integrated Rydberg lifetime $\overline{T}_{\ket{r}}^{\mathrm{opt}} = 3.86/\Omega_{\mathrm{max}}$ \cite{pagano_error_2022}.
Protocols A and B respectively spend 24\% ($\overline{T}_{\ket{r}}^{A} = 4.79/\Omega_{\mathrm{max}}$) and 46\% ($\overline{T}_{\ket{r}}^{B} = 5.64/\Omega_{\mathrm{max}}$) more time in the Rydberg state, than the time-optimal pulses.
This is a worthwhile comparison as the time-optimal pulses also minimize $\overline{T}_{\ket{r}}$ \cite{pagano_error_2022} and with that, the losses due to decay of $\ket{r}$. 
\begin{table}[t]
\centering
\begin{tabular}{l l l} 
 \hline

 \hline\hline
 Principal quantum number $n$ & 61\\ 
 Rabi frequency $\Omega_{\mathrm{max}}$& 2$\pi \times 6.8\;$ & MHz\\ 
vdW coefficient $C_6$ & 2$\pi\times -181$ &
$\mathrm{GHz}\; \mu \mathrm{m}^6$\\
Rydberg lifetime $\tau$ & 96.5 & $\mu$s\\
Interatomic distance $R$& 3.5 & $\mu$m\\

 \hline
\end{tabular}
\caption{Physical parameters considered for $^{88}$Sr}
\label{table:1}
\end{table}
It follows that one trades a reduced lifetime in $\ket{r}$ for robustness -- however, we find that robust pulses with fidelities $F \sim 0.999$ can be obtained over a wide range of control deviations for current $^{88}$Sr experiments. 
\par While two protocols were presented here, a large class of pulses can be generated with the techniques presented in Sec. \ref{QOC}. For example, we observed that pulses with time $T_{\mathrm{opt}} < T < T_B$ -- intermediate between the time-optimal pulse and protocol B -- can be found as well, with both the robustness and $\overline{T}_{\ket{r}}$ intermediate between the two protocols. This presents a way to tailor pulses to a given experimental setup -- thus minimizing infidelity, conditional on the level of deviations present. 
\section{Optimal Rydberg state choice}
\label{whatstate}
In this section we identify the optimal Rydberg state for our setup and show that this can further minimize the infidelity of our gates.
Strontium atoms have long been established for use in optical lattice clocks, and demonstrate accuracy and precision above the Cesium standard \cite{Bothwell_2019, oelker_demonstration_2019}. A technique crucial to the operation of these clocks is \emph{magic wavelength} trapping \cite{doi:10.1126/science.1148259}. This technique aims to eliminate  the differential light shift caused by the trap laser, by identifying wavelengths that cause an identical shift on both clock transition states. For the clock transition introduced in Sec. \ref{atoms}, the magic wavelength $\lambda_m = 813.4$nm \cite{takamoto_optical_2005} is well understood, and often used in experiments. \par
\begin{figure}
     \centering
     \begin{subfigure}[b]{0.41\textwidth}
     \captionsetup{oneside,margin={-0.64cm,0cm}}
         \centering
         \caption{}\includegraphics[width=\textwidth]{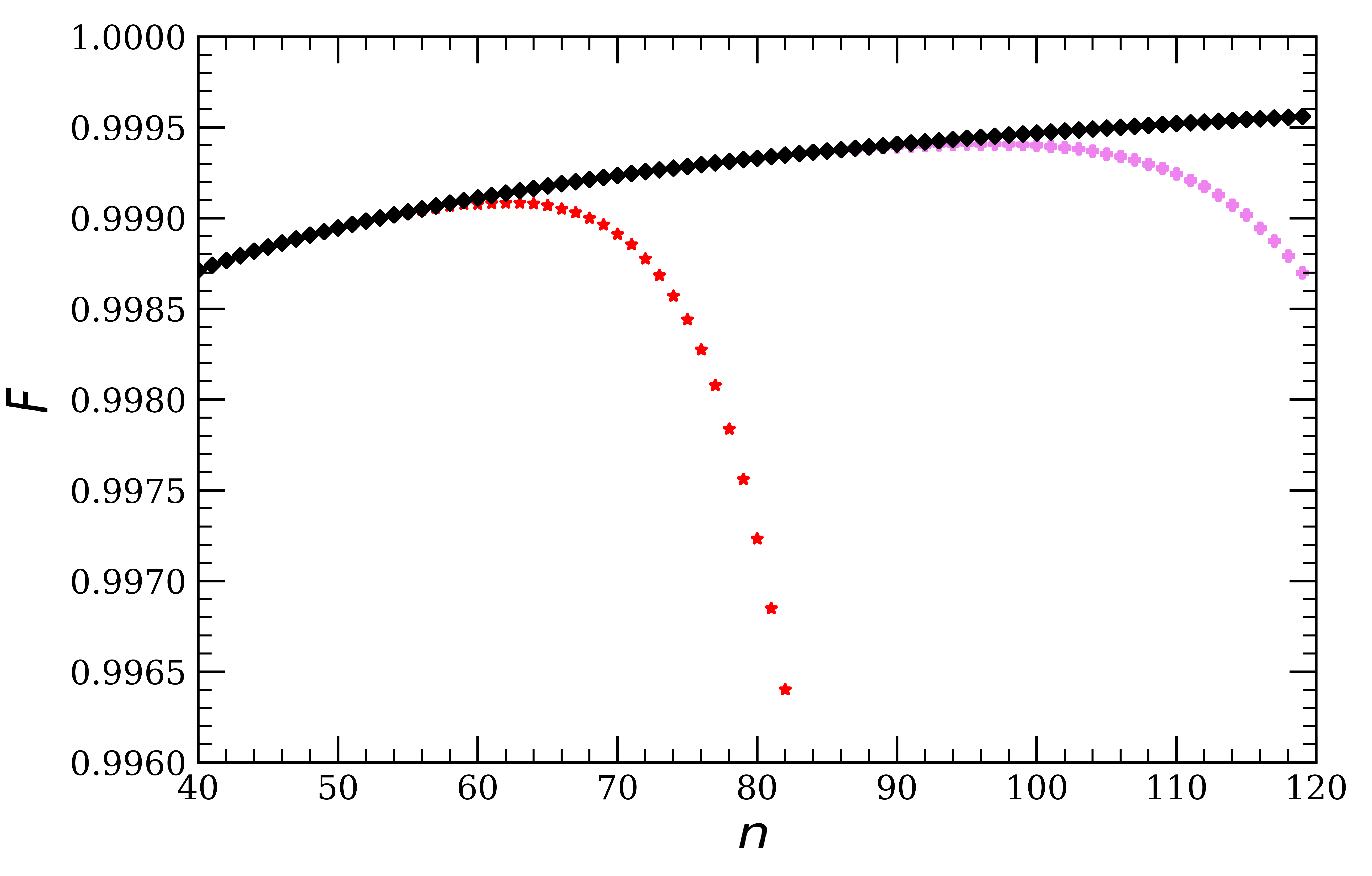}

         \label{fieldsimple}
     \end{subfigure}
     \hfill
     \begin{subfigure}[b]{0.44\textwidth}
         \centering
         \caption{}
         \includegraphics[width=\textwidth]{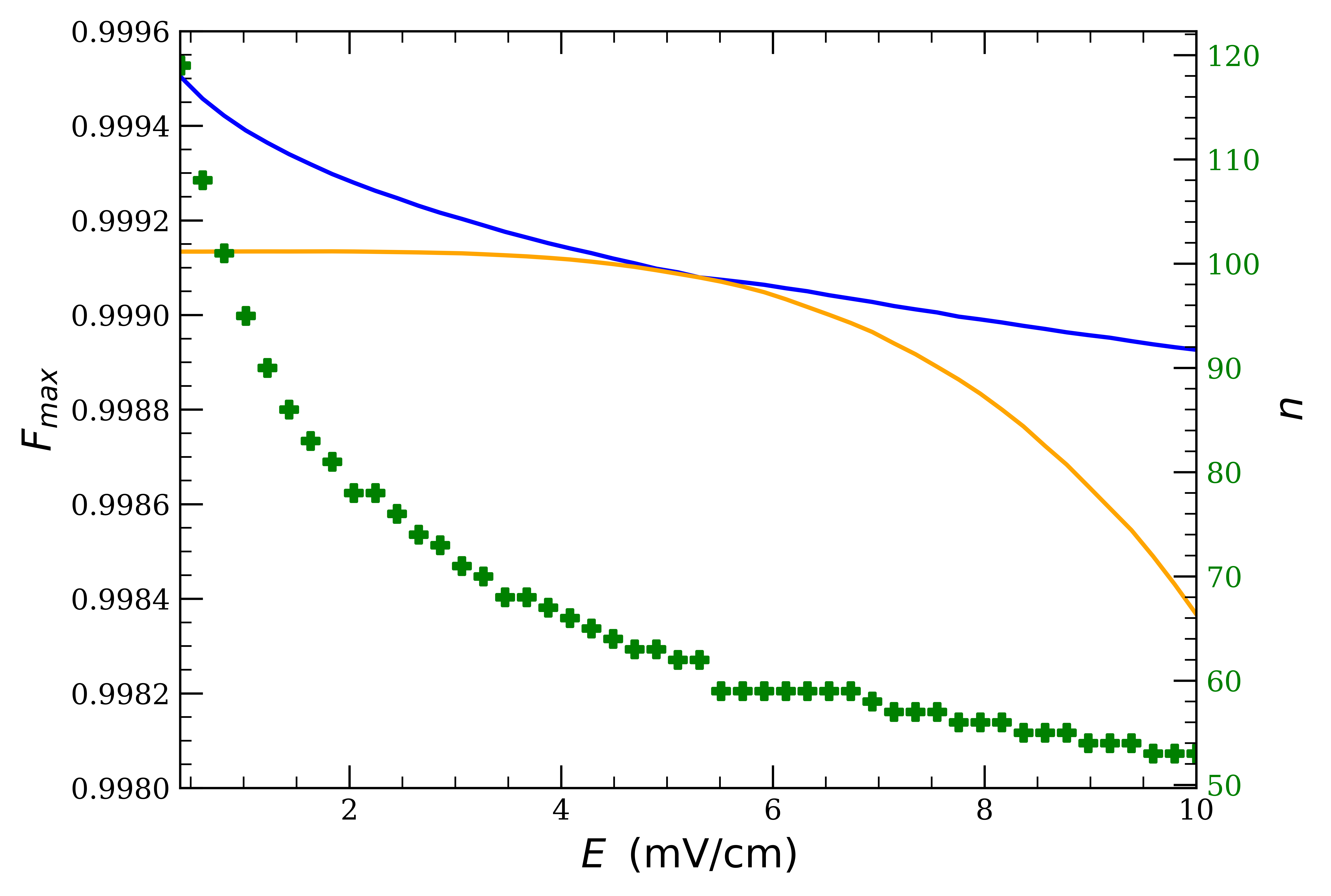}\hspace*{-1.3cm}
         \label{fieldmax}
     \end{subfigure}
     \hfill
\captionsetup{justification=raggedright,singlelinecheck=false}
       \caption{(a) The simulated fidelity $F$ of protocol A (see Fig.~\ref{pulseplot_protocolA}) plotted against the principal quantum number $n$, for E = 0 mV/cm (black, diamond), E = 1 mV/cm (violet, plus) and E = 5 mV/cm (red, stars). As expected from theory, $F$ increases monotonically for the case with no background fields. (b) plots the maximum fidelity (blue), and the $n$ at which this fidelity is obtained (green, plus), against the magnitude of the electric field $E/(\mathrm{mV cm^{-1}}) \in [0.4, 10.0]$. $F$ is also plotted at $n = 61$ (orange).}
        \label{fieldplots}
\end{figure}

Magic traps at this wavelength have also been employed in a quantum information setting \cite{madjarov_entangling_2021, madjarov_high-fidelity_2020}, as this allows for long-lived qubits. For this trap, the Rydberg state is anti-trapped \cite{madjarov_entangling_2021}, and this can lead to loss.\\
One strategy to avoid such loss is to switch off the trap for the duration of the gate, then switch it on for recapture \cite{madjarov_high-fidelity_2020}. The anti-trapping behavior is then mitigated because the spatial wave function will evolve under a free potential instead of a concave Gaussian potential. In this section, we adopt this strategy and thus do not consider trap physics (see Appendix \ref{section:appendixB} for further discussion).\\
In this section, we set again $\Omega_{n=61} = 2\pi \times 6.8$MHz. For a fixed power of the laser, $\Omega \propto (n^*)^{-3/2} $ \cite{Low_2012}, and we use this relation to obtain Rabi frequencies at other values of $n$. The fundamental source of infidelity is the decay from $\ket{r}$ -- characterized by the decay rate $\Gamma$ (see Eqn. \ref{eqn:totallifetime}). Although $\Omega$ decreases with $n$, the ratio
\begin{equation}
\frac{\Omega}{\Gamma} \propto \frac{(n^*)^{-3/2}}{(n^*)^{-3}  + k(n^*)^{-2} },  
\end{equation}
 for a constant $k$, increases with $n$. This ratio captures the two competing time scales of our system -- its increase should thus correspond to an increase in $F$. Figure~\ref{fieldsimple} reports fidelities of protocol A for the range $n \in \{40,\,\dots,\,120\}$ and this increase is indeed observed. For this study, lower bounds on lifetimes of the $\ket{r}$ states were calculated and used -- as detailed in Appendix \ref{section:appendixB}. For $n>120$, the fidelity increases slowly. At $n = 200$, we obtain $F = 0.9997$, compared to $F = 0.9995$ at $n = 100$.\par Stray electric fields in the experiment have been identified as a potential source of infidelity \cite{PhysRevA.97.053803, PhysRevA.87.023423, PhysRevLett.121.123603, saffman_quantum_2016}. As the polarizability of $\ket{r}, \alpha_S \propto (n^*)^7$ \cite{ediss18152}, the associated DC Stark shifts scale in a dramatic manner with $n$, leading to an undesired shift in $\Delta$ -- this is further described in Appendix \ref{section:appendixB}. \\We note that while the shifts due to constant stray fields can be compensated, the drifts in this field result in a shot-to-shot deviation $\delta\Delta$ and pose a significant challenge to the operation of the gate.
 Thus, there are two competing effects that scale with $n$ -- while the lifetime of $\ket{r}$ increases, reducing the infidelity due to decay, the infidelity due to stray fields increases. This competition is illustrated in Fig.~\ref{fieldsimple}.    
Figure~\ref{fieldsimple} also plots fidelities in the presence of residual fields with magnitude $E = 5\mathrm{mV/cm}$ -- this corresponds to significant field cancellations that have been experimentally demonstrated \cite{PhysRevLett.121.123603}. In this case, $F_{\mathrm{max}} = 0.9990$ is obtained at $n = 62$. Looking at another case, with $E = 1\mathrm{mV/cm}$, $F_{\mathrm{max}} = 0.9995$ at $n = 97$.\par Figure~\ref{fieldmax} plots $F_{\mathrm{max}}$ and the optimal $n$ for protocol A, against the residual field strength $E$. Here, we also plot the fidelity of the $n = 61$ state $F_{n=61}$, and note the intersection between the two fidelity curves at $E_{\mathrm{int}} = 5.3$mV/cm. At $E < E_{\mathrm{int}}$, $F_{n=61}$ is limited by the lifetime of $\ket{r}$, and at $E > E_{\mathrm{int}}$ by the DC Stark shifts (see Eq.~\ref{dcstarkshift}).\par  
We conclude that the characterization of residual electric fields in an experiment allows the identification of the optimal Rydberg state for operation. On a NISQ experiment, this allows for $F \sim 0.999$. Higher background field cancellation can achieve $F \sim 0.9995$, and a combination of higher laser power and weaker fields would be required for $F \sim 0.9999$. 
\section{Conclusion}
In this study, we have devised fast robust CZ gates -- with time scales much shorter than the lifetime of the Rydberg states involved, and comparable to the time-optimal gates -- and investigated the associated errors. We found that fidelities $F > 0.999$ are achievable over a wide range of deviations in the laser field coupling $\delta\Omega$ and the detuning $\delta\Delta$. We further identified optimal Rydberg states to maximize fidelities on a NISQ system. \par To the best of our knowledge, this work is the first to carry out a detailed analysis towards the identification of optimal Rydberg states -- we also showed in Sec. \ref{whatstate} that such an analysis can lead to significant gains in two-qubit fidelities, when combined with experimental mitigation of residual electric fields.\par These two results pave the way towards realizing gates with higher fidelities in a $^{88}$Sr experimental setup. The techniques used in our analysis can also be extended to experiments with other neutral atoms.\par  
Future studies could look at identification of optimal Rydberg states for different computation schemes such as VQOC \cite{rvdk}, which would require a careful investigation of the trapping of Rydberg states in the tweezer (over the duration of the quantum circuit). Recent work has been carried out in this direction for $^{88}$Sr atoms \cite{dekeijzer2023recapture}. Optimizing gates to be robust against deviation of the blockade strength is also a worthwhile area of research \cite{jandura_time-optimal_2022}, and the approach presented in our work can be adapted to consider such optimizations.  Additionally, it would be interesting to look at the generation of larger multiqubit gates using these techniques. In the interest of tailoring pulses to specific experiments, rise times of the coupling laser field could be directly incorporated into the optimization framework. 

\par On the experimental side, further techniques to characterize and mitigate background fields could lead to the use of higher-lying Rydberg states, towards even higher fidelities $F \sim 0.9999$ that are conducive to quantum error correction. The verification of our findings on an experimental setup would also greatly benefit the substance of this work -- for instance, through an experiment where laser intensity noise can be controlled, in order to demonstrate the robustness of our pulses. 

\par \emph{During the end of our research, we became aware of related work carried out to investigate robust pulses for two-qubit gates on the Rydberg platform \cite{sven2, fromonteil2022protocols}}.
\begin{acknowledgments}
We acknowledge valuable discussions with Ivo \mbox{Knottnerus}, Alex Urech, Jasper Postema and Sven Jandura. 
Various computational libraries across Python, C++, Fortran and Julia made this work possible -- notable are QuTiP \cite{JOHANSSON20131234} and Altro.jl \cite{Howell-2019-122091} for the variational optimization and simulation. 
ARC3.0 \cite{ROBERTSON2021107814}, FGH \cite{SEATON2002250}, NUMER \cite{SEATON2002254} and the MQDT library from C. Vaillant \cite{Vaillant_2014} were used for the computation of atomic properties. 
This research has received funding from the European Union’s Horizon 2020 research and innovation programme under the Marie Skłodowska-Curie grant agreement number 955479. S.K. and R.K. acknowledge funding from the Dutch Ministry of Economic Affairs and Climate Policy (EZK) as part of the Quantum Delta NL programme, and from the Netherlands Organisation for Scientific Research (NWO) under Grant No. 680-92-18-05.
\end{acknowledgments}
\section*{Data Availability}
The data and code that support the findings of the study -- including the optimized pulses -- are available upon reasonable request to the authors.

\appendix

\section{Calculation of polarizabilities and lifetimes}
\label{section:appendixB}
In this appendix we detail the calculation of atomic properties as introduced in Sec. \ref{atoms}.\par First, we employ quantum defect theory (QDT) \cite{Seaton_1983} towards the calculation of the lifetimes of the $^3\mathrm{S}_1$ Rydberg states of $^{88}$Sr. Our treatment is inspired by the work of Vaillant \emph{et al.} \cite{Vaillant_2014} -- we point the reader to this reference for an in-depth description of QDT towards computing various atomic properties. \par We will use atomic units in this section. $L, S, J$ refer to the total orbital angular momentum, the total spin and the total angular momentum of an atom -- the corresponding lowercase letters $l, s, j$ refer to the momentum of individual electrons. \par The starting point of QDT is the Coulomb approximation: for high lying states, the wavefunction of the active electron is far extended and the potential it experiences can be regarded as a Coulombic potential for large electron distances from the core, $x$. 
QDT then provides a numerical recipe to calculate wavefunctions, and thus to compute the dipole matrix elements -- that capture the atomic physics relevant to our problem, as discussed in the main text. 
\par Dipole matrix elements can be split into angular and the radial dipole matrix elements (RDMEs). A dipole matrix element between two states with quantum numbers $n_1, l_1, m_1$ and $n_2, l_2, m_2$ (for $m$ the magnetic quantum number) respectively can be written as 
\begin{equation}
\bra{n_1l_1m_1}e\mathbf{x}\ket{n_2l_2m_2} = \Braket{l_1m_1 | \frac{\mathbf{x}}{x} | l_2m_2}\bra{n_1l_1}ex\ket{n_2l_2},
\label{eqn:rdme}
\end{equation}
where $x$ is the magnitude of the atomic distance vector of the electron $\mathbf{x}$, $e$ the electron charge. One begins by calculating RDMEs, following which the angular components can be incorporated. \par
For the calculation of the lifetime of the Rydberg state, one must consider both spontaneous emission and black body radiation (BBR) -- at 300K in our setup. For transitions between states $a$ and $b$, the spontaneous decay width is \cite{Vaillant_2014}
\begin{equation}
\Gamma_{a,b}^{sp} = \frac{4\alpha}{3c^2}|\omega_{a,b}|^3 |\bra{\Psi_b}e\mathbf{x}\ket{\Psi_a}|^2,
\end{equation}
where $\omega_{a,b} = E_b - E_a$ is the difference in energies, $\alpha$ is the fine-structure constant and $c$ the speed of light in vacuum. \par
The lifetime due to spontaneous emission for a state $a$, then, is given by
\begin{equation}
\tau_{a}^{sp} = \left(\sum_b \Gamma_{a,b}^{sp}\right)^{-1}.
\end{equation}
The effects of blackbody radiation can be incorporated with the following equation \cite{PhysRevA.23.2397}, 
\begin{equation}
\begin{aligned}
\Gamma_a = \sum_{b, E_a > E_b} \Gamma_{a,b}^{sp}\left( 1 + \frac{1}{\mathrm{e}^{|\omega_{a,b}|/k_BT} - 1} \right) \\ +   \sum_{b, E_a < E_b} \frac{\Gamma_{a,b}^{sp}}{\mathrm{e}^{|\omega_{a,b}|/k_BT} - 1},
\label{eqn:totallifetime} 
\end{aligned}
\end{equation} 
where the first term accounts for the states lying below $a$ and the second term for the higher lying states, for which only BBR contributes to the width. The total lifetime is then $\tau_a = \Gamma_a^{-1}$. \par This treatment is in general valid for alkali atoms, and the situation for $^{88}$Sr is complicated by the presence of two valence electrons in the outermost shell -- for instance, the lifetimes of the $^1S_0$ and $^1D_2$ series in Sr are strongly quenched because of the mixing of doubly-excited low lying states (termed \emph{perturbers}) with the desired singly-excited Rydberg states \cite{Vaillant_2014}. Such perturbers can lead to reduced Rydberg state lifetimes and energy shifts, and thus a careful approach is required in the calculation of physical properties.\par For such a general treatment one needs to consider multichannel quantum defect theory (MQDT), as treated in the main reference, Vaillant \emph{et al.} \cite{Vaillant_2014}. In this context, a channel refers to a specific configuration of the core and the active electron, and perturbers imply a contribution of multiple channels in the atomic properties.\par The $^3\mathrm{S}_1$ series that we use as the Rydberg state is effectively unperturbed \cite{Vaillant_2014, Aymar_1987, Beigang_1982} -- there is a negligible mixture of low-lying doubly excited states -- and thus we do not expect a significant quenching of lifetimes. This also simplifies considerably the theoretical treatment of the lifetimes. \par
Consider the MQDT equation (derived in \cite{Vaillant_2014}) for the natural width of a state $a$,  
\setlength{\belowdisplayskip}{20pt} \setlength{\belowdisplayshortskip}{20pt}
\setlength{\abovedisplayskip}{20pt} \setlength{\abovedisplayshortskip}{20pt}
\begin{equation}
\label{bigeqn}
\begin{aligned}
\Gamma_{a}^{sp} = \frac{4\alpha}{3c^2}\sum_b |\omega_{a,b}|^3 \Bigr[ \sum_{i,j} (-1)^{l_{1,a}^{(i)} + \mathrm{max}(l_{2,a}^{(i)}, l_{2,b}^{(j)}) + S_a^{(i)}} \overline{A}_{ai}\overline{A}_{bj}\\ \times \sqrt{\mathrm{max}(l_{2,a}^{(i)}, l_{2,b}^{(j)})(2L_{b}^{(j)} + 1)(2J_b^{(j)} + 1)(2L_{a}^{(i)} + 1)}\\ \times 
\begin{Bmatrix}
J_b^{(j)} & 1 & J_a^{(i)} \\
L_{a}^{(i)} & S_{a}^{(i)} & L_{b}^{(j)}
\end{Bmatrix}^2 
\begin{Bmatrix}
L_{b}^{(j)} & 1 & L_{a}^{(i)}\\
l_{2a}^{(i)} & l_{1a}^{(i)} & l_{2b}^{(j)}
\end{Bmatrix}^2
R_{b, a}^{(ij)}
 \Bigr]^2
\end{aligned},
\end{equation} 
where $i, j$ are summation indices over the considered channels for states $a, b$ respectively, $\overline{A}_{ai}$ and $\overline{A}_{bj}$ are the respective \emph{channel fractions} (that quantify the mixing of the perturbed states), the curly brackets represent Wigner $6j$ symbols and $R_{b, a}^{(ij)}$ is the RDME corresponding to states $a$ and $b$, in channels $i$ and $j$. \par
\begin{figure}
\includegraphics[width=7.5cm]{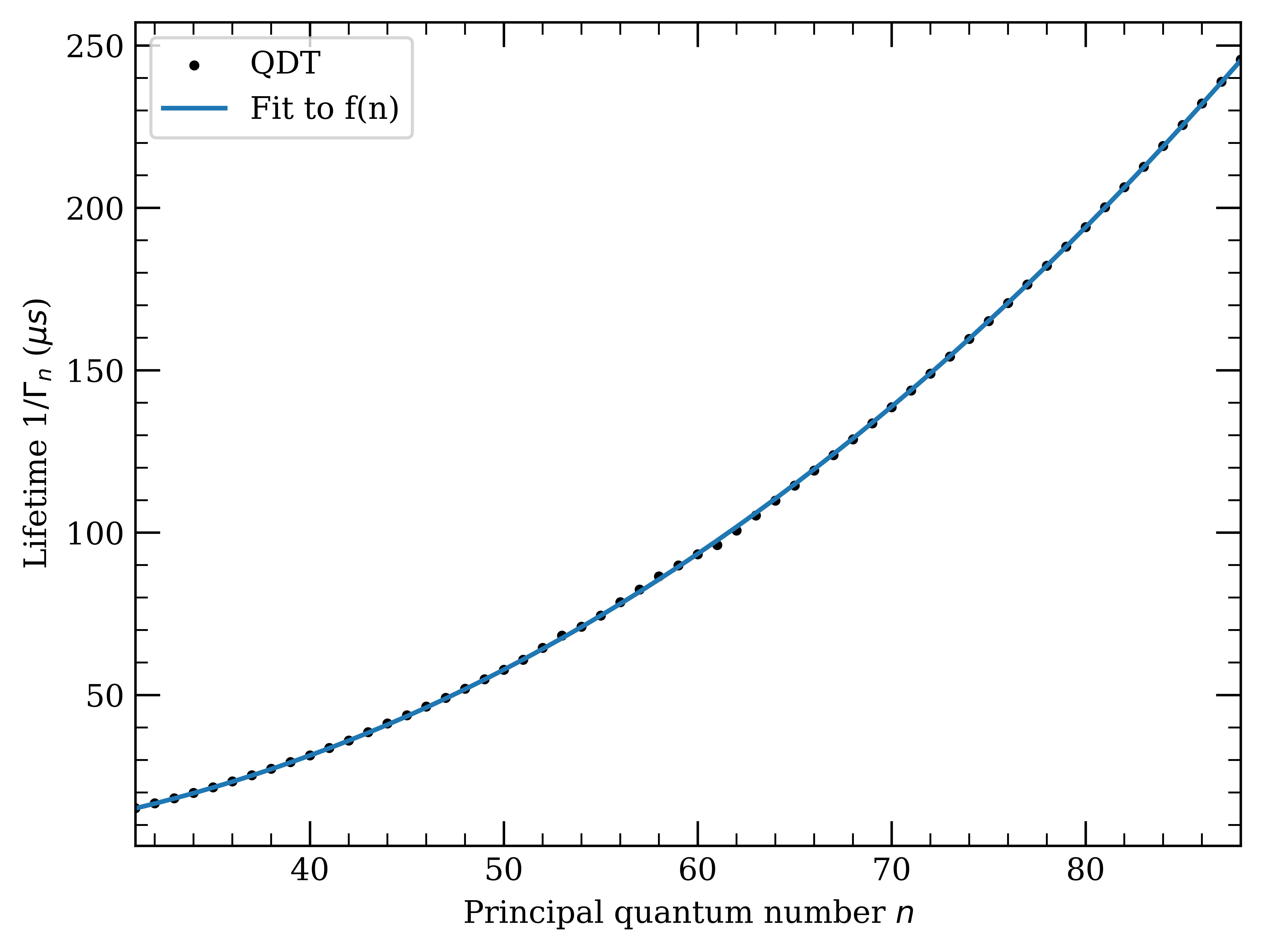}
\caption{Lifetimes (including spontaneous emission and BBR at 300K) for $^{88}$Sr as computed with quantum defect theory (QDT). We find excellent fits to a function of the form $f(n) = 1/(A(n^*)^{-2} + B(n^*)^{-3})$ (for effective principal quantum number $n^*$) as is expected from scaling laws \cite{ediss18152}.}
\label{appendix1}
\end{figure}

The lifetime due to spontaneous emission of state $a$ is then $\tau_a^{sp} = (\Gamma_a^{sp})^{-1}$. Given that the $^3\mathrm{S}_1$ state is effectively unperturbed, it is worth asking whether QDT (the 'single-channel' equivalent of MQDT) can be used. Further, the states that the $^3\mathrm{S}_1$ state decays to -- the $^3\mathrm{P}_{0,1,2}$ states -- are still perturbed, but lack enough experimental data to carry out a full MQDT analysis \cite{Vaillant_2014}.\par We argue that a QDT approach in this case provides a lower bound on the total lifetimes. This is because of three reasons. Firstly, quenching of the lifetimes for divalent atoms arises due to significant radial matrix elements between the doubly-excited states -- as the $^3\mathrm{S}_1$ states do not have these states mixed in, such contributions to the width are negligible. Secondly, the single-excitation wavefunctions can be calculated -- as for Rubidium in \cite{Vaillant_2014}, to excellent agreement with experimental lifetimes -- with QDT. Thirdly, $R_{b, a}$ between a doubly-excited state (mixed in the $^3\mathrm{P}_{0,1,2}$ state) and the singly-excited states is zero. Thus, we use QDT to obtain the lower bounds on the lifetimes in Fig.~\ref{appendix1}-- further studies would have to involve both theory and experiment to obtain MQDT characterizations of the $^3\mathrm{S}_1$ state lifetimes.\par
\begin{figure}
\begin{subfigure}[b]{0.41\textwidth}
         \centering
    \caption{}\includegraphics[width=7.5cm]{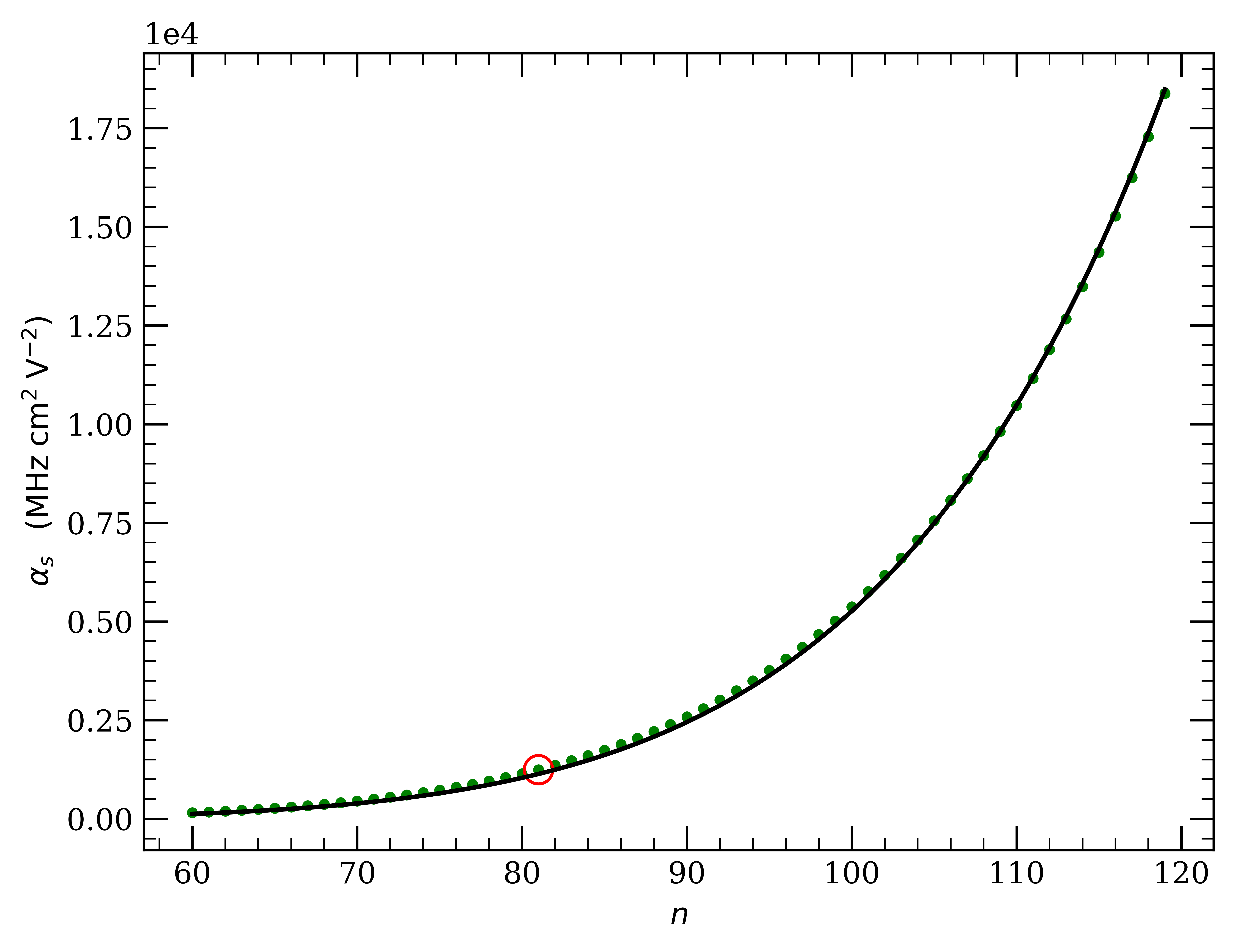}
\end{subfigure}
\begin{subfigure}[b]{0.41\textwidth}
         \centering
         \caption{}
         \includegraphics[width=7.35cm]{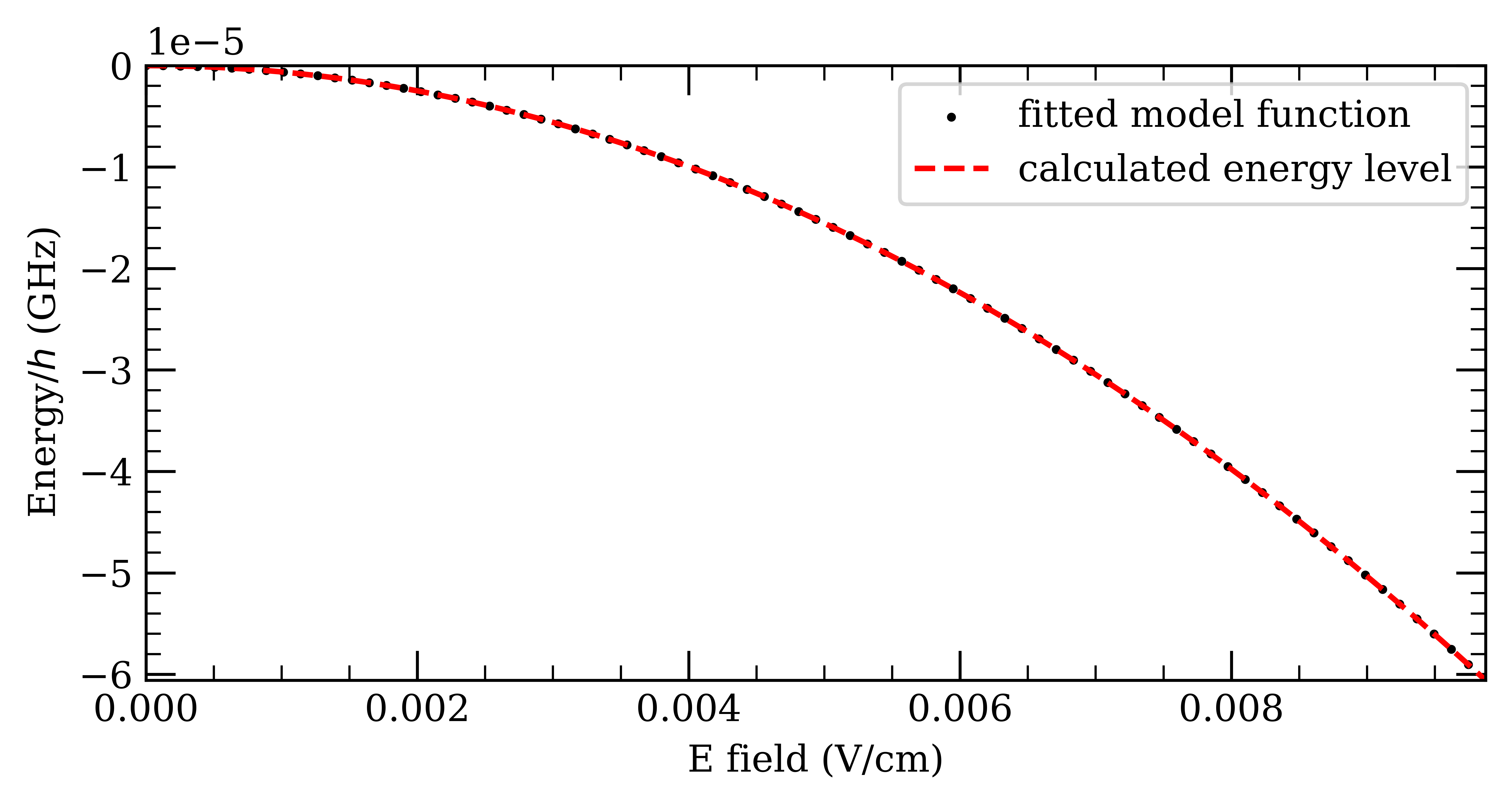}\hspace*{-0.3cm}
         
\end{subfigure}
         
\caption{Scalar polarizabilities $\alpha_S$ for the $5\mathrm{s}n\mathrm{s} \;^3\mathrm{S}_1$ series. (a) The ranges $n \in \{60,\,\dots,\,120\}$ and $|\mathbf{E}|/(\mathrm{mV \; cm^{-1}}) \in [0.0, 10.0]$ are plotted. The values obtained by diagonalization (green, circles) show agreement with the fitting form $f(n) = A(n^*)^7$ (black, line), as expected from theory \cite{madjarov_entangling_2021}. (b) DC Stark shifts for the $n = 81$ state (circled in red on (a)), from which $\alpha_S$ can be deduced -- see Eq. \ref{dcstarkshift}.} 
\label{appendixpolarizability}
\end{figure}
In neutral atom experiments in the NISQ era, stray electric fields are often caused by charges on the glass cell \cite{PhysRevLett.121.123603, PhysRevLett.128.033201}. In Fig.~\ref{appendixpolarizability} we diagonalize the Hamiltonian in the presence of electric fields, to obtain the DC Stark shifts and corresponding polarizabilities for the $5\mathrm{s}n\mathrm{s} ^3\mathrm{S}_1$ Rydberg series. For the considered range of electric fields, we find excellent agreement to models of scalar polarizability -- that is, 
\begin{equation}
\frac{\delta\Delta_{DC}}{2\pi} = -\frac{1}{2}\alpha_S |\mathbf{E}|^2,
\label{dcstarkshift}
\end{equation}
and we extract the polarizability from fits of the DC Stark shift to the electric field ranges in the inset of Fig.~\ref{appendixpolarizability}. The ARC3.0 library \cite{ROBERTSON2021107814} was used for these calculations.\par We now briefly discuss the role of optical dipole traps during the gate operation. Although the trap can be switched off during the pulse (see Sec. \ref{robustsection}), the atom will still expand under the free potential and eventually leave the trap for a long enough duration \cite{propagator}. There are also other challenges when switching off the trap. The blinking on and off of the traps can lead to heating of the atoms and thus heat the entire qubit array. On a sequential gate-based platform, this limits the depth of the gate circuit.
\par

Another possibility is to simply leave the trap on for the Rydberg excitation. It is often assumed that the anti-trapping will then lead to exponential loss of the atoms in time \cite{pagano_error_2022}, however, experimentally it is found that this is not the case \cite{madjarov_high-fidelity_2020,Bluvstein2022,sven2}. We note recent work on the calculation of anti-trapping loss rates for Rydberg states in $^{88}$Sr atoms, that models this scenario \cite{dekeijzer2023recapture}. Further, a recent trapping scheme proposed for $^{88}$Sr \cite{khazali2023subnanometer} -- employing the optical twist of atomic eigenstates -- presents a novel alternative to the optical tweezer traps and holds promise to mitigate errors (including the anti-trapping loss) associated with tweezers, as discussed in this section.

\section{Potential sources of infidelity}
\label{appendix:offresonant}
In this section we elaborate on the error sources introduced in the main text. Further, we discuss additional sources that can be reasonably neglected, as they do not lead to significant infidelities for the parameters considered in the main text, as well as sources that might become important in order to realize higher two-qubit gate fidelities $F \geq 0.9999$.\par When driving the clock transition ($\mathrm{5s^2} \,^1\mathrm{S}_0\leftrightarrow \mathrm{5s5p}^3\mathrm{P}_0$) in $^{88}$Sr, off-resonant scattering can limit the fidelity of operations as the line width of this transition is extremely small \cite{madjarov_entangling_2021}. For this work we calculate off-resonant scattering rates on the $\mathrm{5s5p^3}\mathrm{P}_0$ level and $\mathrm{5s}n\mathrm{s^3}\mathrm{S}_1$ level \cite{steck} using
\begin{equation}
    R_{\mathrm{sc}}=\left(\frac{\Gamma}{2}\right) \frac{\left(I / I_{\mathrm{sat}}\right)}{1+4(\Delta / \Gamma)^2+\left(I / I_{\mathrm{sat}}\right)},
\end{equation}
with $\Gamma$ the decay rate, $I$ the intensity of the laser and $I_{\text{sat}}$ the saturation intesity given by
\begin{equation}
    I_{\mathrm{sat}}=\frac{\hbar \omega^3 \Gamma}{12 \pi c^2},
\end{equation}
where $\hbar$ is the reduced Planck constant, $\omega$ is the frequency of the transition and $c$ is the speed of light \cite{steck}. Decay rates are taken from \cite{decayrates1,decayrates2,decayrates3}.\\
For both the $\mathrm{5s5p^3}\mathrm{P}_0$ and $\mathrm{5s61s}^3\mathrm{S}_1$ levels, off-resonant scattering to all low-lying $^{88}$Sr levels, together with all $\mathrm{5s}n\mathrm{s^3}\mathrm{S}_1$ levels is considered. We found when driving the $\mathrm{5s5p^3P_0}\leftrightarrow \mathrm{5s61s^3}\mathrm{S}_1$ transition at the max Rabi frequency $\Omega_{\mathrm{max}}=2\pi \times 6.8$ MHz, the total off-resonant scattering rate from $\mathrm{5s5p^3}\mathrm{P}_0$ equals 0.14 Hz, where the main contribution is the scattering with $\mathrm{5s5p^3}\mathrm{P}_1$ of 0.11 Hz. For the Rydberg level $\mathrm{5s61s^3}\mathrm{S}_1$, a total off-resonant scattering rate of 0.235 nHz was found for these parameters. On the level of the duration of the pulses, both of these off-resonant scattering processes can be safely ignored. 
\par Now, considering the infidelity contribution due to Doppler shifts, we require the distribution of atomic velocities $\delta v \propto \sqrt{k_B T/m_{\mathrm{Sr}}}$ \cite{PhysRevA.85.042310}, where $k_B$ is the Boltzmann constant, $T$ the atomic temperature and $m_{\mathrm{Sr}}$ the mass of $^{88}$Sr. The corresponding deviation in detuning is given by $\delta\Delta = (2\pi/\lambda_{1r})\times\delta v$, for $\lambda_{1r}$ the wavelength of the Rydberg laser. We take $T = 2.5/\sqrt{10}\,\mu \mathrm{K}$, where the factor of $\sqrt{10}$ arises from adiabatically ramping down the traps by a factor of 10 before carrying out the entangling protocol \cite{madjarov_high-fidelity_2020}. Thus the detuning shift is $\delta\Delta = 4 \times 10^{-3}\,\,\Omega_{\mathrm{max}}$, and the infidelities $(1 - F) \sim 10^{-4}$ for both protocols. \\We further note that while deviations of this order do not cause significant infidelity in $^{88}$Sr, larger $\delta\Delta$ deviations are expected for platforms employing other neutral atom species \cite{mitra_robust_2020} and we expect our protocols to be beneficial in reducing the arising infidelities. \par The interaction strength $V_{\mathrm{int}}$ can also deviate in the presence of stray electric fields \cite{PhysRevA.97.012515, sevincli_2014}. We use the \emph{pairinteraction} library \cite{Weber2017} to compute the shifts of $V_{\mathrm{int}}$ associated with the optimal Rydberg state at the corresponding electric fields in $^{88}$Sr (see Fig. \ref{fieldmax}), and consistently find infidelities $(1 - F) \sim 10^{-6}$. Hence, this error source can be neglected in our study.     

\par In our study we considered $\delta\Omega, \delta\Delta$ to be \emph{shot-to-shot} -- the deviations vary between pulses, but are constant on the time scale of the pulse. We note the current state of the art laser stability in $^{88}$Sr experiments -- Madjarov \emph{et al.} observed intensity deviations $\delta\Omega$ with an RMS deviation $\sigma_{\mathrm{RMS}} = 0.8\%$ \cite{madjarov_high-fidelity_2020}, and the corresponding infidelities can indeed be mitigated with the robust pulses as devised in this work. Although the short time duration of the pulses we devised makes the shot-to-shot character a reasonable assumption, certain time-dependent deviations might also be present in a NISQ experiment -- as a result of components of laser frequency and intensity noise with higher Fourier frequencies. An explicit QOC optimization for such deviations can be carried out with a filter function approach \cite{yang2022quantum, PhysRevA.106.032611}. A detailed analysis of time-dependent deviations will help further the understanding of error sources on the Rydberg platform.\par Finally, although this work considered addressing two atoms with laser pulses, we note that a tweezer array consists of multiple atoms, and infidelities might arise due to laser crosstalk. In particular, there exists a trade-off between crosstalk and laser inhomogeneity at the tweezer sites being addressed \cite{khazali2023electron}, which would further need to be identified for our pulses.
\section{QOC: State vectors and metrics}
\label{qocappendix}
In this appendix, we clarify the variational implementation of the gate dynamics problem with ALTRO \cite{Howell-2019-122091}. The complete augmented state vector is given by
\begin{equation}
\label{formulation_full}
\mathbf{x} = 
\begin{pmatrix}
\Psi, &
\partial_{\Omega}\Psi, &
\partial_{\Delta}\Psi, &
\Omega, &
\Delta, &
\phi, &
P_{\mathrm{tot}}, &
\overline{T}_{\ket{r}}
\end{pmatrix}^{\intercal},
\end{equation}
where $\mathbf{a}^{\intercal}$ refers to the transpose of a vector $\mathbf{a}$.
A reduced vector $\hat{\mathbf{x}}$ was presented in the main text. $\mathbf{x}$ includes all the terms that the controls $\mathbf{c}$ can influence over the pulse duration, and $\hat{\mathbf{x}}$ consists of the terms in $\mathbf{x}$ penalized in the cost function of Eq.~\ref{costfn}. \par We choose a \emph{multi-state transfer} approach \cite{propson_robust_2022} -- the states are evolved individually under the same pulse, which ALTRO then optimizes. The population of a state $\beta$, $P_{\beta} = \bra{\beta}\Psi\ket{\beta}$ and we define the total population as
\begin{equation}
\label{populationtotal}
P_{\mathrm{tot}} = P_{00} + P_{01} + P_{10} + P_{11} = 2P_{01} + P_{11} + 1,
\end{equation}
where $P_{00} = 1$, and $P_{01} = P_{10}$ (see Sec. \ref{QOC}). We set the constraint $P_{\mathrm{tot}} = 4$ at the final time step $N$ to ensure that the computational states are returned to themselves at the end of the protocol. \par The time-integrated probability of being in the Rydberg state $\ket{r}$, with an input computational state $\beta$ \cite{pagano_error_2022}, 
\begin{equation}
T_{\ket{r}}^{\beta} = \int_0^{T_{\mathrm{gate}}}(\Psi^{\beta})^*\, n\,\Psi^{\beta}dt,
\end{equation}
where $n = \ket{r}\bra{r}\otimes I + I\otimes\ket{r}\bra{r}$ and $\Psi^{\beta}$ the time-dependent wavefunction associated with $\beta$ (see Sec. \ref{QOC}). The average integrated Rydberg lifetime, then, 
\begin{equation}
\label{avgtime}
\overline{T}_{\ket{r}} = \frac{1}{3}(2T_{\ket{r}}^{01} + T_{\ket{r}}^{11}),
\end{equation}
where the $\ket{00}$ state is not included, as it does not contribute to infidelity in our protocol.

\newpage

\bibliography{robustcontrol.bib}

\begin{thebibliography}{72}%
\makeatletter
\providecommand \@ifxundefined [1]{%
 \@ifx{#1\undefined}
}%
\providecommand \@ifnum [1]{%
 \ifnum #1\expandafter \@firstoftwo
 \else \expandafter \@secondoftwo
 \fi
}%
\providecommand \@ifx [1]{%
 \ifx #1\expandafter \@firstoftwo
 \else \expandafter \@secondoftwo
 \fi
}%
\providecommand \natexlab [1]{#1}%
\providecommand \enquote  [1]{``#1''}%
\providecommand \bibnamefont  [1]{#1}%
\providecommand \bibfnamefont [1]{#1}%
\providecommand \citenamefont [1]{#1}%
\providecommand \href@noop [0]{\@secondoftwo}%
\providecommand \href [0]{\begingroup \@sanitize@url \@href}%
\providecommand \@href[1]{\@@startlink{#1}\@@href}%
\providecommand \@@href[1]{\endgroup#1\@@endlink}%
\providecommand \@sanitize@url [0]{\catcode `\\12\catcode `\$12\catcode
  `\&12\catcode `\#12\catcode `\^12\catcode `\_12\catcode `\%12\relax}%
\providecommand \@@startlink[1]{}%
\providecommand \@@endlink[0]{}%
\providecommand \url  [0]{\begingroup\@sanitize@url \@url }%
\providecommand \@url [1]{\endgroup\@href {#1}{\urlprefix }}%
\providecommand \urlprefix  [0]{URL }%
\providecommand \Eprint [0]{\href }%
\providecommand \doibase [0]{https://doi.org/}%
\providecommand \selectlanguage [0]{\@gobble}%
\providecommand \bibinfo  [0]{\@secondoftwo}%
\providecommand \bibfield  [0]{\@secondoftwo}%
\providecommand \translation [1]{[#1]}%
\providecommand \BibitemOpen [0]{}%
\providecommand \bibitemStop [0]{}%
\providecommand \bibitemNoStop [0]{.\EOS\space}%
\providecommand \EOS [0]{\spacefactor3000\relax}%
\providecommand \BibitemShut  [1]{\csname bibitem#1\endcsname}%
\let\auto@bib@innerbib\@empty
\bibitem [{\citenamefont {Schymik}\ \emph {et~al.}(2022)\citenamefont
  {Schymik}, \citenamefont {Ximenez}, \citenamefont {Bloch}, \citenamefont
  {Dreon}, \citenamefont {Signoles}, \citenamefont {Nogrette}, \citenamefont
  {Barredo}, \citenamefont {Browaeys},\ and\ \citenamefont
  {Lahaye}}]{PhysRevA.106.022611}%
  \BibitemOpen
  \bibfield  {author} {\bibinfo {author} {\bibfnamefont {K.-N.}\ \bibnamefont
  {Schymik}}, \bibinfo {author} {\bibfnamefont {B.}~\bibnamefont {Ximenez}},
  \bibinfo {author} {\bibfnamefont {E.}~\bibnamefont {Bloch}}, \bibinfo
  {author} {\bibfnamefont {D.}~\bibnamefont {Dreon}}, \bibinfo {author}
  {\bibfnamefont {A.}~\bibnamefont {Signoles}}, \bibinfo {author}
  {\bibfnamefont {F.}~\bibnamefont {Nogrette}}, \bibinfo {author}
  {\bibfnamefont {D.}~\bibnamefont {Barredo}}, \bibinfo {author} {\bibfnamefont
  {A.}~\bibnamefont {Browaeys}},\ and\ \bibinfo {author} {\bibfnamefont
  {T.}~\bibnamefont {Lahaye}},\ }\bibfield  {title} {\bibinfo {title} {{In situ
  equalization of single-atom loading in large-scale optical tweezer arrays}},\
  }\href {https://doi.org/10.1103/PhysRevA.106.022611} {\bibfield  {journal}
  {\bibinfo  {journal} {Phys. Rev. A}\ }\textbf {\bibinfo {volume} {106}},\
  \bibinfo {pages} {022611} (\bibinfo {year} {2022})}\BibitemShut {NoStop}%
\bibitem [{\citenamefont {Ebadi}\ \emph {et~al.}(2021)\citenamefont {Ebadi}
  \emph {et~al.}}]{ebadi_quantum_2021}%
  \BibitemOpen
  \bibfield  {author} {\bibinfo {author} {\bibfnamefont {S.}~\bibnamefont
  {Ebadi}} \emph {et~al.},\ }\bibfield  {title} {\bibinfo {title} {{Quantum
  phases of matter on a 256-atom programmable quantum simulator}},\ }\href
  {https://doi.org/10.1038/s41586-021-03582-4} {\bibfield  {journal} {\bibinfo
  {journal} {Nature (London)}\ }\textbf {\bibinfo {volume} {595}},\ \bibinfo
  {pages} {227} (\bibinfo {year} {2021})}\BibitemShut {NoStop}%
\bibitem [{\citenamefont {Graham}\ \emph {et~al.}(2022)\citenamefont {Graham}
  \emph {et~al.}}]{graham_multi-qubit_2022}%
  \BibitemOpen
  \bibfield  {author} {\bibinfo {author} {\bibfnamefont {T.~M.}\ \bibnamefont
  {Graham}} \emph {et~al.},\ }\bibfield  {title} {\bibinfo {title}
  {{Multi-qubit entanglement and algorithms on a neutral-atom quantum
  computer}},\ }\href {https://doi.org/10.1038/s41586-022-04603-6} {\bibfield
  {journal} {\bibinfo  {journal} {Nature (London)}\ }\textbf {\bibinfo {volume}
  {604}},\ \bibinfo {pages} {457} (\bibinfo {year} {2022})}\BibitemShut
  {NoStop}%
\bibitem [{\citenamefont {Omran}\ \emph {et~al.}(2019)\citenamefont {Omran}
  \emph {et~al.}}]{omranetal2019}%
  \BibitemOpen
  \bibfield  {author} {\bibinfo {author} {\bibfnamefont {A.}~\bibnamefont
  {Omran}} \emph {et~al.},\ }\bibfield  {title} {\bibinfo {title} {{Generation
  and manipulation of Schr\"odinger cat states in Rydberg atom arrays}},\
  }\href {https://doi.org/10.1126/science.aax9743} {\bibfield  {journal}
  {\bibinfo  {journal} {Science}\ }\textbf {\bibinfo {volume} {365}},\ \bibinfo
  {pages} {570} (\bibinfo {year} {2019})}\BibitemShut {NoStop}%
\bibitem [{\citenamefont {Barenco}\ \emph {et~al.}(1995)\citenamefont
  {Barenco}, \citenamefont {Bennett}, \citenamefont {Cleve}, \citenamefont
  {DiVincenzo}, \citenamefont {Margolus}, \citenamefont {Shor}, \citenamefont
  {Sleator}, \citenamefont {Smolin},\ and\ \citenamefont
  {Weinfurter}}]{PhysRevA.52.3457}%
  \BibitemOpen
  \bibfield  {author} {\bibinfo {author} {\bibfnamefont {A.}~\bibnamefont
  {Barenco}}, \bibinfo {author} {\bibfnamefont {C.~H.}\ \bibnamefont
  {Bennett}}, \bibinfo {author} {\bibfnamefont {R.}~\bibnamefont {Cleve}},
  \bibinfo {author} {\bibfnamefont {D.~P.}\ \bibnamefont {DiVincenzo}},
  \bibinfo {author} {\bibfnamefont {N.}~\bibnamefont {Margolus}}, \bibinfo
  {author} {\bibfnamefont {P.}~\bibnamefont {Shor}}, \bibinfo {author}
  {\bibfnamefont {T.}~\bibnamefont {Sleator}}, \bibinfo {author} {\bibfnamefont
  {J.~A.}\ \bibnamefont {Smolin}},\ and\ \bibinfo {author} {\bibfnamefont
  {H.}~\bibnamefont {Weinfurter}},\ }\bibfield  {title} {\bibinfo {title}
  {Elementary gates for quantum computation},\ }\href
  {https://doi.org/10.1103/PhysRevA.52.3457} {\bibfield  {journal} {\bibinfo
  {journal} {Phys. Rev. A}\ }\textbf {\bibinfo {volume} {52}},\ \bibinfo
  {pages} {3457} (\bibinfo {year} {1995})}\BibitemShut {NoStop}%
\bibitem [{\citenamefont {Madjarov}\ \emph {et~al.}(2020)\citenamefont
  {Madjarov}, \citenamefont {Covey}, \citenamefont {Shaw}, \citenamefont
  {Choi}, \citenamefont {Kale}, \citenamefont {Cooper}, \citenamefont
  {Pichler}, \citenamefont {Schkolnik}, \citenamefont {Williams},\ and\
  \citenamefont {Endres}}]{madjarov_high-fidelity_2020}%
  \BibitemOpen
  \bibfield  {author} {\bibinfo {author} {\bibfnamefont {I.~S.}\ \bibnamefont
  {Madjarov}}, \bibinfo {author} {\bibfnamefont {J.~P.}\ \bibnamefont {Covey}},
  \bibinfo {author} {\bibfnamefont {A.~L.}\ \bibnamefont {Shaw}}, \bibinfo
  {author} {\bibfnamefont {J.}~\bibnamefont {Choi}}, \bibinfo {author}
  {\bibfnamefont {A.}~\bibnamefont {Kale}}, \bibinfo {author} {\bibfnamefont
  {A.}~\bibnamefont {Cooper}}, \bibinfo {author} {\bibfnamefont
  {H.}~\bibnamefont {Pichler}}, \bibinfo {author} {\bibfnamefont
  {V.}~\bibnamefont {Schkolnik}}, \bibinfo {author} {\bibfnamefont {J.~R.}\
  \bibnamefont {Williams}},\ and\ \bibinfo {author} {\bibfnamefont
  {M.}~\bibnamefont {Endres}},\ }\bibfield  {title} {\bibinfo {title}
  {{High-fidelity entanglement and detection of alkaline-earth {Rydberg}
  atoms}},\ }\href {https://doi.org/10.1038/s41567-020-0903-z} {\bibfield
  {journal} {\bibinfo  {journal} {Nat. Phys.}\ }\textbf {\bibinfo {volume}
  {16}},\ \bibinfo {pages} {857} (\bibinfo {year} {2020})}\BibitemShut
  {NoStop}%
\bibitem [{\citenamefont {Morgado}\ and\ \citenamefont
  {Whitlock}(2021)}]{shannon2021}%
  \BibitemOpen
  \bibfield  {author} {\bibinfo {author} {\bibfnamefont {M.}~\bibnamefont
  {Morgado}}\ and\ \bibinfo {author} {\bibfnamefont {S.}~\bibnamefont
  {Whitlock}},\ }\bibfield  {title} {\bibinfo {title} {{Quantum simulation and
  computing with Rydberg-interacting qubits}},\ }\href
  {https://doi.org/10.1116/5.0036562} {\bibfield  {journal} {\bibinfo
  {journal} {AVS Quantum Sci.}\ }\textbf {\bibinfo {volume} {3}},\ \bibinfo
  {pages} {023501} (\bibinfo {year} {2021})}\BibitemShut {NoStop}%
\bibitem [{\citenamefont {Theis}\ \emph {et~al.}(2016)\citenamefont {Theis},
  \citenamefont {Motzoi}, \citenamefont {Wilhelm},\ and\ \citenamefont
  {Saffman}}]{PhysRevA.94.032306}%
  \BibitemOpen
  \bibfield  {author} {\bibinfo {author} {\bibfnamefont {L.~S.}\ \bibnamefont
  {Theis}}, \bibinfo {author} {\bibfnamefont {F.}~\bibnamefont {Motzoi}},
  \bibinfo {author} {\bibfnamefont {F.~K.}\ \bibnamefont {Wilhelm}},\ and\
  \bibinfo {author} {\bibfnamefont {M.}~\bibnamefont {Saffman}},\ }\bibfield
  {title} {\bibinfo {title} {{High-fidelity Rydberg-blockade entangling gate
  using shaped, analytic pulses}},\ }\href
  {https://doi.org/10.1103/PhysRevA.94.032306} {\bibfield  {journal} {\bibinfo
  {journal} {Phys. Rev. A}\ }\textbf {\bibinfo {volume} {94}},\ \bibinfo
  {pages} {032306} (\bibinfo {year} {2016})}\BibitemShut {NoStop}%
\bibitem [{\citenamefont {Devitt}\ \emph {et~al.}(2013)\citenamefont {Devitt},
  \citenamefont {Munro},\ and\ \citenamefont {Nemoto}}]{Devitt_2013}%
  \BibitemOpen
  \bibfield  {author} {\bibinfo {author} {\bibfnamefont {S.~J.}\ \bibnamefont
  {Devitt}}, \bibinfo {author} {\bibfnamefont {W.~J.}\ \bibnamefont {Munro}},\
  and\ \bibinfo {author} {\bibfnamefont {K.}~\bibnamefont {Nemoto}},\
  }\bibfield  {title} {\bibinfo {title} {Quantum error correction for
  beginners},\ }\href {https://doi.org/10.1088/0034-4885/76/7/076001}
  {\bibfield  {journal} {\bibinfo  {journal} {Rep. Prog. Phys.}\ }\textbf
  {\bibinfo {volume} {76}},\ \bibinfo {pages} {076001} (\bibinfo {year}
  {2013})}\BibitemShut {NoStop}%
\bibitem [{\citenamefont {Preskill}(2018)}]{Preskill2018quantumcomputingin}%
  \BibitemOpen
  \bibfield  {author} {\bibinfo {author} {\bibfnamefont {J.}~\bibnamefont
  {Preskill}},\ }\bibfield  {title} {\bibinfo {title} {Quantum {C}omputing in
  the {NISQ} era and beyond},\ }\href
  {https://doi.org/10.22331/q-2018-08-06-79} {\bibfield  {journal} {\bibinfo
  {journal} {{Quantum}}\ }\textbf {\bibinfo {volume} {2}},\ \bibinfo {pages}
  {79} (\bibinfo {year} {2018})}\BibitemShut {NoStop}%
\bibitem [{\citenamefont {Jaksch}\ \emph {et~al.}(2000)\citenamefont {Jaksch},
  \citenamefont {Cirac}, \citenamefont {Zoller}, \citenamefont {Rolston},
  \citenamefont {C\^ot\'e},\ and\ \citenamefont {Lukin}}]{PhysRevLett.85.2208}%
  \BibitemOpen
  \bibfield  {author} {\bibinfo {author} {\bibfnamefont {D.}~\bibnamefont
  {Jaksch}}, \bibinfo {author} {\bibfnamefont {J.~I.}\ \bibnamefont {Cirac}},
  \bibinfo {author} {\bibfnamefont {P.}~\bibnamefont {Zoller}}, \bibinfo
  {author} {\bibfnamefont {S.~L.}\ \bibnamefont {Rolston}}, \bibinfo {author}
  {\bibfnamefont {R.}~\bibnamefont {C\^ot\'e}},\ and\ \bibinfo {author}
  {\bibfnamefont {M.~D.}\ \bibnamefont {Lukin}},\ }\bibfield  {title} {\bibinfo
  {title} {{Fast Quantum Gates for Neutral Atoms}},\ }\href
  {https://doi.org/10.1103/PhysRevLett.85.2208} {\bibfield  {journal} {\bibinfo
   {journal} {Phys. Rev. Lett.}\ }\textbf {\bibinfo {volume} {85}},\ \bibinfo
  {pages} {2208} (\bibinfo {year} {2000})}\BibitemShut {NoStop}%
\bibitem [{\citenamefont {Mitra}\ \emph {et~al.}(2020)\citenamefont {Mitra},
  \citenamefont {Martin}, \citenamefont {Biedermann}, \citenamefont {Marino},
  \citenamefont {Poggi},\ and\ \citenamefont {Deutsch}}]{mitra_robust_2020}%
  \BibitemOpen
  \bibfield  {author} {\bibinfo {author} {\bibfnamefont {A.}~\bibnamefont
  {Mitra}}, \bibinfo {author} {\bibfnamefont {M.~J.}\ \bibnamefont {Martin}},
  \bibinfo {author} {\bibfnamefont {G.~W.}\ \bibnamefont {Biedermann}},
  \bibinfo {author} {\bibfnamefont {A.~M.}\ \bibnamefont {Marino}}, \bibinfo
  {author} {\bibfnamefont {P.~M.}\ \bibnamefont {Poggi}},\ and\ \bibinfo
  {author} {\bibfnamefont {I.~H.}\ \bibnamefont {Deutsch}},\ }\bibfield
  {title} {\bibinfo {title} {{Robust M{\o}lmer-S{\o}rensen gate for neutral
  atoms using rapid adiabatic {Rydberg} dressing}},\ }\href
  {https://doi.org/10.1103/PhysRevA.101.030301} {\bibfield  {journal} {\bibinfo
   {journal} {Phys. Rev. A}\ }\textbf {\bibinfo {volume} {101}},\ \bibinfo
  {pages} {030301(R)} (\bibinfo {year} {2020})}\BibitemShut {NoStop}%
\bibitem [{\citenamefont {Saffman}(2016)}]{saffman_quantum_2016}%
  \BibitemOpen
  \bibfield  {author} {\bibinfo {author} {\bibfnamefont {M.}~\bibnamefont
  {Saffman}},\ }\bibfield  {title} {\bibinfo {title} {{Quantum computing with
  atomic qubits and {Rydberg} interactions: progress and challenges}},\ }\href
  {https://doi.org/10.1088/0953-4075/49/20/202001} {\bibfield  {journal}
  {\bibinfo  {journal} {J. Phys. B}\ }\textbf {\bibinfo {volume} {49}},\
  \bibinfo {pages} {202001} (\bibinfo {year} {2016})}\BibitemShut {NoStop}%
\bibitem [{\citenamefont {Jandura}\ and\ \citenamefont
  {Pupillo}(2022)}]{jandura_time-optimal_2022}%
  \BibitemOpen
  \bibfield  {author} {\bibinfo {author} {\bibfnamefont {S.}~\bibnamefont
  {Jandura}}\ and\ \bibinfo {author} {\bibfnamefont {G.}~\bibnamefont
  {Pupillo}},\ }\bibfield  {title} {\bibinfo {title} {{Time-{Optimal} {Two}-
  and {Three}-{Qubit} {Gates} for {Rydberg} {Atoms}}},\ }\href
  {https://doi.org/10.22331/q-2022-05-13-712} {\bibfield  {journal} {\bibinfo
  {journal} {Quantum}\ }\textbf {\bibinfo {volume} {6}},\ \bibinfo {pages}
  {712} (\bibinfo {year} {2022})}\BibitemShut {NoStop}%
\bibitem [{\citenamefont {Pagano}\ \emph {et~al.}(2022)\citenamefont {Pagano},
  \citenamefont {Weber}, \citenamefont {Jaschke}, \citenamefont {Pfau},
  \citenamefont {Meinert}, \citenamefont {Montangero},\ and\ \citenamefont
  {Büchler}}]{pagano_error_2022}%
  \BibitemOpen
  \bibfield  {author} {\bibinfo {author} {\bibfnamefont {A.}~\bibnamefont
  {Pagano}}, \bibinfo {author} {\bibfnamefont {S.}~\bibnamefont {Weber}},
  \bibinfo {author} {\bibfnamefont {D.}~\bibnamefont {Jaschke}}, \bibinfo
  {author} {\bibfnamefont {T.}~\bibnamefont {Pfau}}, \bibinfo {author}
  {\bibfnamefont {F.}~\bibnamefont {Meinert}}, \bibinfo {author} {\bibfnamefont
  {S.}~\bibnamefont {Montangero}},\ and\ \bibinfo {author} {\bibfnamefont
  {H.~P.}\ \bibnamefont {Büchler}},\ }\bibfield  {title} {\bibinfo {title}
  {{Error budgeting for a controlled-phase gate with strontium-88 {Rydberg}
  atoms}},\ }\href {https://doi.org/10.1103/PhysRevResearch.4.033019}
  {\bibfield  {journal} {\bibinfo  {journal} {Phys. Rev. Research}\ }\textbf
  {\bibinfo {volume} {4}},\ \bibinfo {pages} {033019} (\bibinfo {year}
  {2022})}\BibitemShut {NoStop}%
\bibitem [{\citenamefont {Koch}(2016)}]{Koch_2016}%
  \BibitemOpen
  \bibfield  {author} {\bibinfo {author} {\bibfnamefont {C.~P.}\ \bibnamefont
  {Koch}},\ }\bibfield  {title} {\bibinfo {title} {Controlling open quantum
  systems: tools, achievements, and limitations},\ }\href
  {https://doi.org/10.1088/0953-8984/28/21/213001} {\bibfield  {journal}
  {\bibinfo  {journal} {J. Phys. Cond. Mat.}\ }\textbf {\bibinfo {volume}
  {28}},\ \bibinfo {pages} {213001} (\bibinfo {year} {2016})}\BibitemShut
  {NoStop}%
\bibitem [{\citenamefont {Heeres}\ \emph {et~al.}(2017)\citenamefont {Heeres},
  \citenamefont {Reinhold}, \citenamefont {Ofek}, \citenamefont {Frunzio},
  \citenamefont {Jiang}, \citenamefont {Devoret},\ and\ \citenamefont
  {Schoelkopf}}]{heeres_implementing_2017}%
  \BibitemOpen
  \bibfield  {author} {\bibinfo {author} {\bibfnamefont {R.~W.}\ \bibnamefont
  {Heeres}}, \bibinfo {author} {\bibfnamefont {P.}~\bibnamefont {Reinhold}},
  \bibinfo {author} {\bibfnamefont {N.}~\bibnamefont {Ofek}}, \bibinfo {author}
  {\bibfnamefont {L.}~\bibnamefont {Frunzio}}, \bibinfo {author} {\bibfnamefont
  {L.}~\bibnamefont {Jiang}}, \bibinfo {author} {\bibfnamefont {M.~H.}\
  \bibnamefont {Devoret}},\ and\ \bibinfo {author} {\bibfnamefont {R.~J.}\
  \bibnamefont {Schoelkopf}},\ }\bibfield  {title} {\bibinfo {title}
  {{Implementing a universal gate set on a logical qubit encoded in an
  oscillator}},\ }\href {https://doi.org/10.1038/s41467-017-00045-1} {\bibfield
   {journal} {\bibinfo  {journal} {Nat. Commun.}\ }\textbf {\bibinfo {volume}
  {8}},\ \bibinfo {pages} {94} (\bibinfo {year} {2017})}\BibitemShut {NoStop}%
\bibitem [{\citenamefont {Matekole}\ \emph {et~al.}(2022)\citenamefont
  {Matekole}, \citenamefont {Fang},\ and\ \citenamefont {Lin}}]{9835639}%
  \BibitemOpen
  \bibfield  {author} {\bibinfo {author} {\bibfnamefont {E.~S.}\ \bibnamefont
  {Matekole}}, \bibinfo {author} {\bibfnamefont {Y.-L.~L.}\ \bibnamefont
  {Fang}},\ and\ \bibinfo {author} {\bibfnamefont {M.}~\bibnamefont {Lin}},\
  }\bibfield  {title} {\bibinfo {title} {{Methods and Results for Quantum
  Optimal Pulse Control on Superconducting Qubit Systems}},\ }in\ \href
  {https://doi.org/10.1109/IPDPSW55747.2022.00102} {\emph {\bibinfo {booktitle}
  {2022 IEEE International Parallel and Distributed Processing Symposium
  Workshops (IPDPSW)}}}\ (\bibinfo {year} {2022})\ pp.\ \bibinfo {pages}
  {600--606}\BibitemShut {NoStop}%
\bibitem [{\citenamefont {Werninghaus}\ \emph {et~al.}(2021)\citenamefont
  {Werninghaus}, \citenamefont {Egger}, \citenamefont {Roy}, \citenamefont
  {Machnes}, \citenamefont {Wilhelm},\ and\ \citenamefont
  {Filipp}}]{werninghaus_leakage_2021}%
  \BibitemOpen
  \bibfield  {author} {\bibinfo {author} {\bibfnamefont {M.}~\bibnamefont
  {Werninghaus}}, \bibinfo {author} {\bibfnamefont {D.~J.}\ \bibnamefont
  {Egger}}, \bibinfo {author} {\bibfnamefont {F.}~\bibnamefont {Roy}}, \bibinfo
  {author} {\bibfnamefont {S.}~\bibnamefont {Machnes}}, \bibinfo {author}
  {\bibfnamefont {F.~K.}\ \bibnamefont {Wilhelm}},\ and\ \bibinfo {author}
  {\bibfnamefont {S.}~\bibnamefont {Filipp}},\ }\bibfield  {title} {\bibinfo
  {title} {Leakage reduction in fast superconducting qubit gates via optimal
  control},\ }\href {https://doi.org/10.1038/s41534-020-00346-2} {\bibfield
  {journal} {\bibinfo  {journal} {npj Quantum Information}\ }\textbf {\bibinfo
  {volume} {7}},\ \bibinfo {pages} {1} (\bibinfo {year} {2021})}\BibitemShut
  {NoStop}%
\bibitem [{\citenamefont {Wu}\ \emph {et~al.}(2020)\citenamefont {Wu},
  \citenamefont {Tomarken}, \citenamefont {Petersson}, \citenamefont
  {Martinez}, \citenamefont {Rosen},\ and\ \citenamefont
  {DuBois}}]{PhysRevLett.125.170502}%
  \BibitemOpen
  \bibfield  {author} {\bibinfo {author} {\bibfnamefont {X.}~\bibnamefont
  {Wu}}, \bibinfo {author} {\bibfnamefont {S.~L.}\ \bibnamefont {Tomarken}},
  \bibinfo {author} {\bibfnamefont {N.~A.}\ \bibnamefont {Petersson}}, \bibinfo
  {author} {\bibfnamefont {L.~A.}\ \bibnamefont {Martinez}}, \bibinfo {author}
  {\bibfnamefont {Y.~J.}\ \bibnamefont {Rosen}},\ and\ \bibinfo {author}
  {\bibfnamefont {J.~L.}\ \bibnamefont {DuBois}},\ }\bibfield  {title}
  {\bibinfo {title} {{High-Fidelity Software-Defined Quantum Logic on a
  Superconducting Qudit}},\ }\href
  {https://doi.org/10.1103/PhysRevLett.125.170502} {\bibfield  {journal}
  {\bibinfo  {journal} {Phys. Rev. Lett.}\ }\textbf {\bibinfo {volume} {125}},\
  \bibinfo {pages} {170502} (\bibinfo {year} {2020})}\BibitemShut {NoStop}%
\bibitem [{\citenamefont {Larrouy}\ \emph {et~al.}(2020)\citenamefont
  {Larrouy}, \citenamefont {Patsch}, \citenamefont {Richaud}, \citenamefont
  {Raimond}, \citenamefont {Brune}, \citenamefont {Koch},\ and\ \citenamefont
  {Gleyzes}}]{PhysRevX.10.021058}%
  \BibitemOpen
  \bibfield  {author} {\bibinfo {author} {\bibfnamefont {A.}~\bibnamefont
  {Larrouy}}, \bibinfo {author} {\bibfnamefont {S.}~\bibnamefont {Patsch}},
  \bibinfo {author} {\bibfnamefont {R.}~\bibnamefont {Richaud}}, \bibinfo
  {author} {\bibfnamefont {J.-M.}\ \bibnamefont {Raimond}}, \bibinfo {author}
  {\bibfnamefont {M.}~\bibnamefont {Brune}}, \bibinfo {author} {\bibfnamefont
  {C.~P.}\ \bibnamefont {Koch}},\ and\ \bibinfo {author} {\bibfnamefont
  {S.}~\bibnamefont {Gleyzes}},\ }\bibfield  {title} {\bibinfo {title} {{Fast
  Navigation in a Large Hilbert Space Using Quantum Optimal Control}},\ }\href
  {https://doi.org/10.1103/PhysRevX.10.021058} {\bibfield  {journal} {\bibinfo
  {journal} {Phys. Rev. X}\ }\textbf {\bibinfo {volume} {10}},\ \bibinfo
  {pages} {021058} (\bibinfo {year} {2020})}\BibitemShut {NoStop}%
\bibitem [{\citenamefont {Weggemans}\ \emph {et~al.}(2022)\citenamefont
  {Weggemans}, \citenamefont {Urech}, \citenamefont {Rausch}, \citenamefont
  {Spreeuw}, \citenamefont {Boucherie}, \citenamefont {Schreck}, \citenamefont
  {Schoutens}, \citenamefont {Min{\'{a}}{\v{r}}},\ and\ \citenamefont
  {Speelman}}]{Weggemans2022solvingcorrelation}%
  \BibitemOpen
  \bibfield  {author} {\bibinfo {author} {\bibfnamefont {J.~R.}\ \bibnamefont
  {Weggemans}}, \bibinfo {author} {\bibfnamefont {A.}~\bibnamefont {Urech}},
  \bibinfo {author} {\bibfnamefont {A.}~\bibnamefont {Rausch}}, \bibinfo
  {author} {\bibfnamefont {R.}~\bibnamefont {Spreeuw}}, \bibinfo {author}
  {\bibfnamefont {R.}~\bibnamefont {Boucherie}}, \bibinfo {author}
  {\bibfnamefont {F.}~\bibnamefont {Schreck}}, \bibinfo {author} {\bibfnamefont
  {K.}~\bibnamefont {Schoutens}}, \bibinfo {author} {\bibfnamefont
  {J.}~\bibnamefont {Min{\'{a}}{\v{r}}}},\ and\ \bibinfo {author}
  {\bibfnamefont {F.}~\bibnamefont {Speelman}},\ }\bibfield  {title} {\bibinfo
  {title} {{Solving correlation clustering with {QAOA} and a {R}ydberg qudit
  system: a full-stack approach}},\ }\href
  {https://doi.org/10.22331/q-2022-04-13-687} {\bibfield  {journal} {\bibinfo
  {journal} {{Quantum}}\ }\textbf {\bibinfo {volume} {6}},\ \bibinfo {pages}
  {687} (\bibinfo {year} {2022})}\BibitemShut {NoStop}%
\bibitem [{\citenamefont {Daley}\ \emph {et~al.}(2008)\citenamefont {Daley},
  \citenamefont {Boyd}, \citenamefont {Ye},\ and\ \citenamefont
  {Zoller}}]{PhysRevLett.101.170504}%
  \BibitemOpen
  \bibfield  {author} {\bibinfo {author} {\bibfnamefont {A.~J.}\ \bibnamefont
  {Daley}}, \bibinfo {author} {\bibfnamefont {M.~M.}\ \bibnamefont {Boyd}},
  \bibinfo {author} {\bibfnamefont {J.}~\bibnamefont {Ye}},\ and\ \bibinfo
  {author} {\bibfnamefont {P.}~\bibnamefont {Zoller}},\ }\bibfield  {title}
  {\bibinfo {title} {{Quantum Computing with Alkaline-Earth-Metal Atoms}},\
  }\href {https://doi.org/10.1103/PhysRevLett.101.170504} {\bibfield  {journal}
  {\bibinfo  {journal} {Phys. Rev. Lett.}\ }\textbf {\bibinfo {volume} {101}},\
  \bibinfo {pages} {170504} (\bibinfo {year} {2008})}\BibitemShut {NoStop}%
\bibitem [{\citenamefont {Chen}\ \emph {et~al.}(2022)\citenamefont {Chen},
  \citenamefont {Li}, \citenamefont {Huie}, \citenamefont {Zhao}, \citenamefont
  {Vetter}, \citenamefont {Greene},\ and\ \citenamefont
  {Covey}}]{PhysRevA.105.052438}%
  \BibitemOpen
  \bibfield  {author} {\bibinfo {author} {\bibfnamefont {N.}~\bibnamefont
  {Chen}}, \bibinfo {author} {\bibfnamefont {L.}~\bibnamefont {Li}}, \bibinfo
  {author} {\bibfnamefont {W.}~\bibnamefont {Huie}}, \bibinfo {author}
  {\bibfnamefont {M.}~\bibnamefont {Zhao}}, \bibinfo {author} {\bibfnamefont
  {I.}~\bibnamefont {Vetter}}, \bibinfo {author} {\bibfnamefont {C.~H.}\
  \bibnamefont {Greene}},\ and\ \bibinfo {author} {\bibfnamefont {J.~P.}\
  \bibnamefont {Covey}},\ }\bibfield  {title} {\bibinfo {title} {{Analyzing the
  Rydberg-based optical-metastable-ground architecture for $^{171}\mathrm{Yb}$
  nuclear spins}},\ }\href {https://doi.org/10.1103/PhysRevA.105.052438}
  {\bibfield  {journal} {\bibinfo  {journal} {Phys. Rev. A}\ }\textbf {\bibinfo
  {volume} {105}},\ \bibinfo {pages} {052438} (\bibinfo {year}
  {2022})}\BibitemShut {NoStop}%
\bibitem [{\citenamefont {Madjarov}(2021)}]{madjarov_entangling_2021}%
  \BibitemOpen
  \bibfield  {author} {\bibinfo {author} {\bibfnamefont {I.~S.}\ \bibnamefont
  {Madjarov}},\ }\emph {\bibinfo {title} {{Entangling, {Controlling}, and
  {Detecting} {Individual} {Strontium} {Atoms} in {Optical} {Tweezer}
  {Arrays}}}},\ \href {https://doi.org/10.7907/d1em-dt34} {Ph.D. thesis},\
  \bibinfo  {school} {California Institute of Technology} (\bibinfo {year}
  {2021})\BibitemShut {NoStop}%
\bibitem [{\citenamefont {Cong}\ \emph {et~al.}(2022)\citenamefont {Cong},
  \citenamefont {Levine}, \citenamefont {Keesling}, \citenamefont {Bluvstein},
  \citenamefont {Wang},\ and\ \citenamefont {Lukin}}]{PhysRevX.12.021049}%
  \BibitemOpen
  \bibfield  {author} {\bibinfo {author} {\bibfnamefont {I.}~\bibnamefont
  {Cong}}, \bibinfo {author} {\bibfnamefont {H.}~\bibnamefont {Levine}},
  \bibinfo {author} {\bibfnamefont {A.}~\bibnamefont {Keesling}}, \bibinfo
  {author} {\bibfnamefont {D.}~\bibnamefont {Bluvstein}}, \bibinfo {author}
  {\bibfnamefont {S.-T.}\ \bibnamefont {Wang}},\ and\ \bibinfo {author}
  {\bibfnamefont {M.~D.}\ \bibnamefont {Lukin}},\ }\bibfield  {title} {\bibinfo
  {title} {{Hardware-Efficient, Fault-Tolerant Quantum Computation with Rydberg
  Atoms}},\ }\href {https://doi.org/10.1103/PhysRevX.12.021049} {\bibfield
  {journal} {\bibinfo  {journal} {Phys. Rev. X}\ }\textbf {\bibinfo {volume}
  {12}},\ \bibinfo {pages} {021049} (\bibinfo {year} {2022})}\BibitemShut
  {NoStop}%
\bibitem [{\citenamefont {Levine}\ \emph {et~al.}(2018)\citenamefont {Levine},
  \citenamefont {Keesling}, \citenamefont {Omran}, \citenamefont {Bernien},
  \citenamefont {Schwartz}, \citenamefont {Zibrov}, \citenamefont {Endres},
  \citenamefont {Greiner}, \citenamefont {Vuleti\ifmmode~\acute{c}\else
  \'{c}\fi{}},\ and\ \citenamefont {Lukin}}]{PhysRevLett.121.123603}%
  \BibitemOpen
  \bibfield  {author} {\bibinfo {author} {\bibfnamefont {H.}~\bibnamefont
  {Levine}}, \bibinfo {author} {\bibfnamefont {A.}~\bibnamefont {Keesling}},
  \bibinfo {author} {\bibfnamefont {A.}~\bibnamefont {Omran}}, \bibinfo
  {author} {\bibfnamefont {H.}~\bibnamefont {Bernien}}, \bibinfo {author}
  {\bibfnamefont {S.}~\bibnamefont {Schwartz}}, \bibinfo {author}
  {\bibfnamefont {A.~S.}\ \bibnamefont {Zibrov}}, \bibinfo {author}
  {\bibfnamefont {M.}~\bibnamefont {Endres}}, \bibinfo {author} {\bibfnamefont
  {M.}~\bibnamefont {Greiner}}, \bibinfo {author} {\bibfnamefont
  {V.}~\bibnamefont {Vuleti\ifmmode~\acute{c}\else \'{c}\fi{}}},\ and\ \bibinfo
  {author} {\bibfnamefont {M.~D.}\ \bibnamefont {Lukin}},\ }\bibfield  {title}
  {\bibinfo {title} {{High-Fidelity Control and Entanglement of {R}ydberg-Atom
  Qubits}},\ }\href {https://doi.org/10.1103/PhysRevLett.121.123603} {\bibfield
   {journal} {\bibinfo  {journal} {Phys. Rev. Lett.}\ }\textbf {\bibinfo
  {volume} {121}},\ \bibinfo {pages} {123603} (\bibinfo {year}
  {2018})}\BibitemShut {NoStop}%
\bibitem [{\citenamefont {Seaton}(1983)}]{Seaton_1983}%
  \BibitemOpen
  \bibfield  {author} {\bibinfo {author} {\bibfnamefont {M.~J.}\ \bibnamefont
  {Seaton}},\ }\bibfield  {title} {\bibinfo {title} {Quantum defect theory},\
  }\href {https://doi.org/10.1088/0034-4885/46/2/002} {\bibfield  {journal}
  {\bibinfo  {journal} {Rep. Prog. Phys.}\ }\textbf {\bibinfo {volume} {46}},\
  \bibinfo {pages} {167} (\bibinfo {year} {1983})}\BibitemShut {NoStop}%
\bibitem [{\citenamefont {{Gounand, F.}}(1979)}]{gounand1979}%
  \BibitemOpen
  \bibfield  {author} {\bibinfo {author} {\bibnamefont {{Gounand, F.}}},\
  }\bibfield  {title} {\bibinfo {title} {{Calculation of radial matrix elements
  and radiative lifetimes for highly excited states of alkali atoms using the
  Coulomb approximation}},\ }\href
  {https://doi.org/10.1051/jphys:01979004005045700} {\bibfield  {journal}
  {\bibinfo  {journal} {J. Phys. France}\ }\textbf {\bibinfo {volume} {40}},\
  \bibinfo {pages} {457} (\bibinfo {year} {1979})}\BibitemShut {NoStop}%
\bibitem [{\citenamefont {Beterov}\ \emph {et~al.}(2009)\citenamefont
  {Beterov}, \citenamefont {Ryabtsev}, \citenamefont {Tretyakov},\ and\
  \citenamefont {Entin}}]{PhysRevA.79.052504}%
  \BibitemOpen
  \bibfield  {author} {\bibinfo {author} {\bibfnamefont {I.~I.}\ \bibnamefont
  {Beterov}}, \bibinfo {author} {\bibfnamefont {I.~I.}\ \bibnamefont
  {Ryabtsev}}, \bibinfo {author} {\bibfnamefont {D.~B.}\ \bibnamefont
  {Tretyakov}},\ and\ \bibinfo {author} {\bibfnamefont {V.~M.}\ \bibnamefont
  {Entin}},\ }\bibfield  {title} {\bibinfo {title} {{Quasiclassical
  calculations of blackbody-radiation-induced depopulation rates and effective
  lifetimes of Rydberg $nS$, $nP$, and $nD$ alkali-metal atoms with
  $n\ensuremath{\le}80$}},\ }\href {https://doi.org/10.1103/PhysRevA.79.052504}
  {\bibfield  {journal} {\bibinfo  {journal} {Phys. Rev. A}\ }\textbf {\bibinfo
  {volume} {79}},\ \bibinfo {pages} {052504} (\bibinfo {year}
  {2009})}\BibitemShut {NoStop}%
\bibitem [{\citenamefont {Schauss}(2015)}]{ediss18152}%
  \BibitemOpen
  \bibfield  {author} {\bibinfo {author} {\bibfnamefont {P.}~\bibnamefont
  {Schauss}},\ }\emph {\bibinfo {title} {{High-resolution imaging of ordering
  in Rydberg many-body systems}}},\ \href
  {http://nbn-resolving.de/urn:nbn:de:bvb:19-181524} {Ph.D. thesis},\ \bibinfo
  {school} {Ludwig-Maximilians-Universit{\"a}t M{\"u}nchen} (\bibinfo {year}
  {2015})\BibitemShut {NoStop}%
\bibitem [{\citenamefont {Vaillant}\ \emph {et~al.}(2014)\citenamefont
  {Vaillant}, \citenamefont {Jones},\ and\ \citenamefont
  {Potvliege}}]{Vaillant_2014}%
  \BibitemOpen
  \bibfield  {author} {\bibinfo {author} {\bibfnamefont {C.~L.}\ \bibnamefont
  {Vaillant}}, \bibinfo {author} {\bibfnamefont {M.~P.~A.}\ \bibnamefont
  {Jones}},\ and\ \bibinfo {author} {\bibfnamefont {R.~M.}\ \bibnamefont
  {Potvliege}},\ }\bibfield  {title} {\bibinfo {title} {{Multichannel quantum
  defect theory of strontium bound Rydberg states}},\ }\href
  {https://doi.org/10.1088/0953-4075/47/15/155001} {\bibfield  {journal}
  {\bibinfo  {journal} {J. Phys. B}\ }\textbf {\bibinfo {volume} {47}},\
  \bibinfo {pages} {155001} (\bibinfo {year} {2014})}\BibitemShut {NoStop}%
\bibitem [{\citenamefont {Petrosyan}\ \emph {et~al.}(2017)\citenamefont
  {Petrosyan}, \citenamefont {Motzoi}, \citenamefont {Saffman},\ and\
  \citenamefont {M\o{}lmer}}]{PhysRevA.96.042306}%
  \BibitemOpen
  \bibfield  {author} {\bibinfo {author} {\bibfnamefont {D.}~\bibnamefont
  {Petrosyan}}, \bibinfo {author} {\bibfnamefont {F.}~\bibnamefont {Motzoi}},
  \bibinfo {author} {\bibfnamefont {M.}~\bibnamefont {Saffman}},\ and\ \bibinfo
  {author} {\bibfnamefont {K.}~\bibnamefont {M\o{}lmer}},\ }\bibfield  {title}
  {\bibinfo {title} {{High-fidelity Rydberg quantum gate via a two-atom dark
  state}},\ }\href {https://doi.org/10.1103/PhysRevA.96.042306} {\bibfield
  {journal} {\bibinfo  {journal} {Phys. Rev. A}\ }\textbf {\bibinfo {volume}
  {96}},\ \bibinfo {pages} {042306} (\bibinfo {year} {2017})}\BibitemShut
  {NoStop}%
\bibitem [{\citenamefont {Saffman}\ \emph {et~al.}(2010)\citenamefont
  {Saffman}, \citenamefont {Walker},\ and\ \citenamefont
  {M\o{}lmer}}]{RevModPhys.82.2313}%
  \BibitemOpen
  \bibfield  {author} {\bibinfo {author} {\bibfnamefont {M.}~\bibnamefont
  {Saffman}}, \bibinfo {author} {\bibfnamefont {T.~G.}\ \bibnamefont
  {Walker}},\ and\ \bibinfo {author} {\bibfnamefont {K.}~\bibnamefont
  {M\o{}lmer}},\ }\bibfield  {title} {\bibinfo {title} {{Quantum information
  with Rydberg atoms}},\ }\href {https://doi.org/10.1103/RevModPhys.82.2313}
  {\bibfield  {journal} {\bibinfo  {journal} {Rev. Mod. Phys.}\ }\textbf
  {\bibinfo {volume} {82}},\ \bibinfo {pages} {2313} (\bibinfo {year}
  {2010})}\BibitemShut {NoStop}%
\bibitem [{\citenamefont {Saffman}\ \emph {et~al.}(2020)\citenamefont
  {Saffman}, \citenamefont {Beterov}, \citenamefont {Dalal}, \citenamefont
  {Páez},\ and\ \citenamefont {Sanders}}]{saffman_symmetric_2020}%
  \BibitemOpen
  \bibfield  {author} {\bibinfo {author} {\bibfnamefont {M.}~\bibnamefont
  {Saffman}}, \bibinfo {author} {\bibfnamefont {I.~I.}\ \bibnamefont
  {Beterov}}, \bibinfo {author} {\bibfnamefont {A.}~\bibnamefont {Dalal}},
  \bibinfo {author} {\bibfnamefont {E.~J.}\ \bibnamefont {Páez}},\ and\
  \bibinfo {author} {\bibfnamefont {B.~C.}\ \bibnamefont {Sanders}},\
  }\bibfield  {title} {\bibinfo {title} {{Symmetric {Rydberg}
  controlled-\${Z}\$ gates with adiabatic pulses}},\ }\href
  {https://doi.org/10.1103/PhysRevA.101.062309} {\bibfield  {journal} {\bibinfo
   {journal} {Phys. Rev. A}\ }\textbf {\bibinfo {volume} {101}},\ \bibinfo
  {pages} {062309} (\bibinfo {year} {2020})}\BibitemShut {NoStop}%
\bibitem [{\citenamefont {Goerz}\ \emph {et~al.}(2014)\citenamefont {Goerz},
  \citenamefont {Halperin}, \citenamefont {Aytac}, \citenamefont {Koch},\ and\
  \citenamefont {Whaley}}]{PhysRevA.90.032329}%
  \BibitemOpen
  \bibfield  {author} {\bibinfo {author} {\bibfnamefont {M.~H.}\ \bibnamefont
  {Goerz}}, \bibinfo {author} {\bibfnamefont {E.~J.}\ \bibnamefont {Halperin}},
  \bibinfo {author} {\bibfnamefont {J.~M.}\ \bibnamefont {Aytac}}, \bibinfo
  {author} {\bibfnamefont {C.~P.}\ \bibnamefont {Koch}},\ and\ \bibinfo
  {author} {\bibfnamefont {K.~B.}\ \bibnamefont {Whaley}},\ }\bibfield  {title}
  {\bibinfo {title} {{Robustness of high-fidelity Rydberg gates with
  single-site addressability}},\ }\href
  {https://doi.org/10.1103/PhysRevA.90.032329} {\bibfield  {journal} {\bibinfo
  {journal} {Phys. Rev. A}\ }\textbf {\bibinfo {volume} {90}},\ \bibinfo
  {pages} {032329} (\bibinfo {year} {2014})}\BibitemShut {NoStop}%
\bibitem [{\citenamefont {Guo}\ \emph {et~al.}(2020)\citenamefont {Guo},
  \citenamefont {Yan}, \citenamefont {Zhang}, \citenamefont {Su},\ and\
  \citenamefont {Li}}]{PhysRevA.102.042607}%
  \BibitemOpen
  \bibfield  {author} {\bibinfo {author} {\bibfnamefont {C.-Y.}\ \bibnamefont
  {Guo}}, \bibinfo {author} {\bibfnamefont {L.-L.}\ \bibnamefont {Yan}},
  \bibinfo {author} {\bibfnamefont {S.}~\bibnamefont {Zhang}}, \bibinfo
  {author} {\bibfnamefont {S.-L.}\ \bibnamefont {Su}},\ and\ \bibinfo {author}
  {\bibfnamefont {W.}~\bibnamefont {Li}},\ }\bibfield  {title} {\bibinfo
  {title} {{Optimized geometric quantum computation with a mesoscopic ensemble
  of Rydberg atoms}},\ }\href {https://doi.org/10.1103/PhysRevA.102.042607}
  {\bibfield  {journal} {\bibinfo  {journal} {Phys. Rev. A}\ }\textbf {\bibinfo
  {volume} {102}},\ \bibinfo {pages} {042607} (\bibinfo {year}
  {2020})}\BibitemShut {NoStop}%
\bibitem [{\citenamefont {Howell}\ \emph {et~al.}(2019)\citenamefont {Howell},
  \citenamefont {Jackson},\ and\ \citenamefont
  {Manchester}}]{Howell-2019-122091}%
  \BibitemOpen
  \bibfield  {author} {\bibinfo {author} {\bibfnamefont {T.}~\bibnamefont
  {Howell}}, \bibinfo {author} {\bibfnamefont {B.}~\bibnamefont {Jackson}},\
  and\ \bibinfo {author} {\bibfnamefont {Z.}~\bibnamefont {Manchester}},\
  }\bibfield  {title} {\bibinfo {title} {{ALTRO: A Fast Solver for Constrained
  Trajectory Optimization}},\ }in\ \href
  {https://www.ri.cmu.edu/wp-content/uploads/2020/06/altro-iros.pdf} {\emph
  {\bibinfo {booktitle} {Proceedings of (IROS) IEEE/RSJ International
  Conference on Intelligent Robots and Systems}}}\ (\bibinfo {year} {2019})\
  pp.\ \bibinfo {pages} {7674 -- 7679}\BibitemShut {NoStop}%
\bibitem [{\citenamefont {Propson}\ \emph {et~al.}(2022)\citenamefont
  {Propson}, \citenamefont {Jackson}, \citenamefont {Koch}, \citenamefont
  {Manchester},\ and\ \citenamefont {Schuster}}]{propson_robust_2022}%
  \BibitemOpen
  \bibfield  {author} {\bibinfo {author} {\bibfnamefont {T.}~\bibnamefont
  {Propson}}, \bibinfo {author} {\bibfnamefont {B.~E.}\ \bibnamefont
  {Jackson}}, \bibinfo {author} {\bibfnamefont {J.}~\bibnamefont {Koch}},
  \bibinfo {author} {\bibfnamefont {Z.}~\bibnamefont {Manchester}},\ and\
  \bibinfo {author} {\bibfnamefont {D.~I.}\ \bibnamefont {Schuster}},\
  }\bibfield  {title} {\bibinfo {title} {{Robust {Quantum} {Optimal} {Control}
  with {Trajectory} {Optimization}}},\ }\href
  {https://doi.org/10.1103/PhysRevApplied.17.014036} {\bibfield  {journal}
  {\bibinfo  {journal} {Phys. Rev. Applied}\ }\textbf {\bibinfo {volume}
  {17}},\ \bibinfo {pages} {014036} (\bibinfo {year} {2022})}\BibitemShut
  {NoStop}%
\bibitem [{\citenamefont {de~Keijzer}\ \emph {et~al.}(2022)\citenamefont
  {de~Keijzer}, \citenamefont {Tse},\ and\ \citenamefont {Kokkelmans}}]{rvdk}%
  \BibitemOpen
  \bibfield  {author} {\bibinfo {author} {\bibfnamefont {R.}~\bibnamefont
  {de~Keijzer}}, \bibinfo {author} {\bibfnamefont {O.}~\bibnamefont {Tse}},\
  and\ \bibinfo {author} {\bibfnamefont {S.}~\bibnamefont {Kokkelmans}},\
  }\href {https://doi.org/10.48550/ARXIV.2202.08908} {\bibinfo {title} {{Pulse
  based Variational Quantum Optimal Control for hybrid quantum computing}}}
  (\bibinfo {year} {2022}),\ \Eprint {https://arxiv.org/abs/2202.08908}
  {arXiv:2202.08908 [quant-ph]} \BibitemShut {NoStop}%
\bibitem [{\citenamefont {Robertson}\ \emph {et~al.}(2021)\citenamefont
  {Robertson}, \citenamefont {Šibalić}, \citenamefont {Potvliege},\ and\
  \citenamefont {Jones}}]{ROBERTSON2021107814}%
  \BibitemOpen
  \bibfield  {author} {\bibinfo {author} {\bibfnamefont {E.}~\bibnamefont
  {Robertson}}, \bibinfo {author} {\bibfnamefont {N.}~\bibnamefont
  {Šibalić}}, \bibinfo {author} {\bibfnamefont {R.}~\bibnamefont
  {Potvliege}},\ and\ \bibinfo {author} {\bibfnamefont {M.}~\bibnamefont
  {Jones}},\ }\bibfield  {title} {\bibinfo {title} {{ARC 3.0: An expanded
  Python toolbox for atomic physics calculations}},\ }\href
  {https://doi.org/https://doi.org/10.1016/j.cpc.2020.107814} {\bibfield
  {journal} {\bibinfo  {journal} {Comput. Phys. Commun.}\ }\textbf {\bibinfo
  {volume} {261}},\ \bibinfo {pages} {107814} (\bibinfo {year}
  {2021})}\BibitemShut {NoStop}%
\bibitem [{\citenamefont {Bothwell}\ \emph {et~al.}(2019)\citenamefont
  {Bothwell}, \citenamefont {Kedar}, \citenamefont {Oelker}, \citenamefont
  {Robinson}, \citenamefont {Bromley}, \citenamefont {Tew}, \citenamefont
  {Ye},\ and\ \citenamefont {Kennedy}}]{Bothwell_2019}%
  \BibitemOpen
  \bibfield  {author} {\bibinfo {author} {\bibfnamefont {T.}~\bibnamefont
  {Bothwell}}, \bibinfo {author} {\bibfnamefont {D.}~\bibnamefont {Kedar}},
  \bibinfo {author} {\bibfnamefont {E.}~\bibnamefont {Oelker}}, \bibinfo
  {author} {\bibfnamefont {J.~M.}\ \bibnamefont {Robinson}}, \bibinfo {author}
  {\bibfnamefont {S.~L.}\ \bibnamefont {Bromley}}, \bibinfo {author}
  {\bibfnamefont {W.~L.}\ \bibnamefont {Tew}}, \bibinfo {author} {\bibfnamefont
  {J.}~\bibnamefont {Ye}},\ and\ \bibinfo {author} {\bibfnamefont {C.~J.}\
  \bibnamefont {Kennedy}},\ }\bibfield  {title} {\bibinfo {title} {{JILA SrI
  optical lattice clock with uncertainty of $2.0 \times 10^{-18}$}},\ }\href
  {https://doi.org/10.1088/1681-7575/ab4089} {\bibfield  {journal} {\bibinfo
  {journal} {Metrologia}\ }\textbf {\bibinfo {volume} {56}},\ \bibinfo {pages}
  {065004} (\bibinfo {year} {2019})}\BibitemShut {NoStop}%
\bibitem [{\citenamefont {Oelker}\ \emph {et~al.}(2019)\citenamefont {Oelker}
  \emph {et~al.}}]{oelker_demonstration_2019}%
  \BibitemOpen
  \bibfield  {author} {\bibinfo {author} {\bibfnamefont {E.}~\bibnamefont
  {Oelker}} \emph {et~al.},\ }\bibfield  {title} {\bibinfo {title}
  {{Demonstration of $4.8 \times 10^{−17}$ stability at 1 s for two
  independent optical clocks}},\ }\href
  {https://doi.org/10.1038/s41566-019-0493-4} {\bibfield  {journal} {\bibinfo
  {journal} {Nat. Photon.}\ }\textbf {\bibinfo {volume} {13}},\ \bibinfo
  {pages} {714} (\bibinfo {year} {2019})}\BibitemShut {NoStop}%
\bibitem [{\citenamefont {Ye}\ \emph {et~al.}(2008)\citenamefont {Ye},
  \citenamefont {Kimble},\ and\ \citenamefont
  {Katori}}]{doi:10.1126/science.1148259}%
  \BibitemOpen
  \bibfield  {author} {\bibinfo {author} {\bibfnamefont {J.}~\bibnamefont
  {Ye}}, \bibinfo {author} {\bibfnamefont {H.~J.}\ \bibnamefont {Kimble}},\
  and\ \bibinfo {author} {\bibfnamefont {H.}~\bibnamefont {Katori}},\
  }\bibfield  {title} {\bibinfo {title} {{Quantum State Engineering and
  Precision Metrology Using State-Insensitive Light Traps}},\ }\href
  {https://doi.org/10.1126/science.1148259} {\bibfield  {journal} {\bibinfo
  {journal} {Science}\ }\textbf {\bibinfo {volume} {320}},\ \bibinfo {pages}
  {1734} (\bibinfo {year} {2008})}\BibitemShut {NoStop}%
\bibitem [{\citenamefont {Takamoto}\ \emph {et~al.}(2005)\citenamefont
  {Takamoto}, \citenamefont {Hong}, \citenamefont {Higashi},\ and\
  \citenamefont {Katori}}]{takamoto_optical_2005}%
  \BibitemOpen
  \bibfield  {author} {\bibinfo {author} {\bibfnamefont {M.}~\bibnamefont
  {Takamoto}}, \bibinfo {author} {\bibfnamefont {F.-L.}\ \bibnamefont {Hong}},
  \bibinfo {author} {\bibfnamefont {R.}~\bibnamefont {Higashi}},\ and\ \bibinfo
  {author} {\bibfnamefont {H.}~\bibnamefont {Katori}},\ }\bibfield  {title}
  {\bibinfo {title} {An optical lattice clock},\ }\href
  {https://doi.org/10.1038/nature03541} {\bibfield  {journal} {\bibinfo
  {journal} {Nature (London)}\ }\textbf {\bibinfo {volume} {435}},\ \bibinfo
  {pages} {321} (\bibinfo {year} {2005})}\BibitemShut {NoStop}%
\bibitem [{\citenamefont {Löw}\ \emph {et~al.}(2012)\citenamefont {Löw},
  \citenamefont {Weimer}, \citenamefont {Nipper}, \citenamefont {Balewski},
  \citenamefont {Butscher}, \citenamefont {Büchler},\ and\ \citenamefont
  {Pfau}}]{Low_2012}%
  \BibitemOpen
  \bibfield  {author} {\bibinfo {author} {\bibfnamefont {R.}~\bibnamefont
  {Löw}}, \bibinfo {author} {\bibfnamefont {H.}~\bibnamefont {Weimer}},
  \bibinfo {author} {\bibfnamefont {J.}~\bibnamefont {Nipper}}, \bibinfo
  {author} {\bibfnamefont {J.~B.}\ \bibnamefont {Balewski}}, \bibinfo {author}
  {\bibfnamefont {B.}~\bibnamefont {Butscher}}, \bibinfo {author}
  {\bibfnamefont {H.~P.}\ \bibnamefont {Büchler}},\ and\ \bibinfo {author}
  {\bibfnamefont {T.}~\bibnamefont {Pfau}},\ }\bibfield  {title} {\bibinfo
  {title} {{An experimental and theoretical guide to strongly interacting
  Rydberg gases}},\ }\href {https://doi.org/10.1088/0953-4075/45/11/113001}
  {\bibfield  {journal} {\bibinfo  {journal} {J. Phys. B}\ }\textbf {\bibinfo
  {volume} {45}},\ \bibinfo {pages} {113001} (\bibinfo {year}
  {2012})}\BibitemShut {NoStop}%
\bibitem [{\citenamefont {de~L\'es\'eleuc}\ \emph {et~al.}(2018)\citenamefont
  {de~L\'es\'eleuc}, \citenamefont {Barredo}, \citenamefont {Lienhard},
  \citenamefont {Browaeys},\ and\ \citenamefont {Lahaye}}]{PhysRevA.97.053803}%
  \BibitemOpen
  \bibfield  {author} {\bibinfo {author} {\bibfnamefont {S.}~\bibnamefont
  {de~L\'es\'eleuc}}, \bibinfo {author} {\bibfnamefont {D.}~\bibnamefont
  {Barredo}}, \bibinfo {author} {\bibfnamefont {V.}~\bibnamefont {Lienhard}},
  \bibinfo {author} {\bibfnamefont {A.}~\bibnamefont {Browaeys}},\ and\
  \bibinfo {author} {\bibfnamefont {T.}~\bibnamefont {Lahaye}},\ }\bibfield
  {title} {\bibinfo {title} {{Analysis of imperfections in the coherent optical
  excitation of single atoms to Rydberg states}},\ }\href
  {https://doi.org/10.1103/PhysRevA.97.053803} {\bibfield  {journal} {\bibinfo
  {journal} {Phys. Rev. A}\ }\textbf {\bibinfo {volume} {97}},\ \bibinfo
  {pages} {053803} (\bibinfo {year} {2018})}\BibitemShut {NoStop}%
\bibitem [{\citenamefont {Jones}\ \emph {et~al.}(2013)\citenamefont {Jones},
  \citenamefont {Carter},\ and\ \citenamefont {Martin}}]{PhysRevA.87.023423}%
  \BibitemOpen
  \bibfield  {author} {\bibinfo {author} {\bibfnamefont {L.~A.}\ \bibnamefont
  {Jones}}, \bibinfo {author} {\bibfnamefont {J.~D.}\ \bibnamefont {Carter}},\
  and\ \bibinfo {author} {\bibfnamefont {J.~D.~D.}\ \bibnamefont {Martin}},\
  }\bibfield  {title} {\bibinfo {title} {{Rydberg atoms with a reduced
  sensitivity to dc and low-frequency electric fields}},\ }\href
  {https://doi.org/10.1103/PhysRevA.87.023423} {\bibfield  {journal} {\bibinfo
  {journal} {Phys. Rev. A}\ }\textbf {\bibinfo {volume} {87}},\ \bibinfo
  {pages} {023423} (\bibinfo {year} {2013})}\BibitemShut {NoStop}%
\bibitem [{\citenamefont {de~Keijzer}\ \emph {et~al.}(2023)\citenamefont
  {de~Keijzer}, \citenamefont {Tse},\ and\ \citenamefont
  {Kokkelmans}}]{dekeijzer2023recapture}%
  \BibitemOpen
  \bibfield  {author} {\bibinfo {author} {\bibfnamefont {R.}~\bibnamefont
  {de~Keijzer}}, \bibinfo {author} {\bibfnamefont {O.}~\bibnamefont {Tse}},\
  and\ \bibinfo {author} {\bibfnamefont {S.}~\bibnamefont {Kokkelmans}},\
  }\href@noop {} {\bibinfo {title} {{Recapture Probability for anti-trapped
  Rydberg states in optical tweezers}}} (\bibinfo {year} {2023}),\ \Eprint
  {https://arxiv.org/abs/2303.08783} {arXiv:2303.08783 [quant-ph]} \BibitemShut
  {NoStop}%
\bibitem [{\citenamefont {Jandura}\ \emph {et~al.}(2022)\citenamefont
  {Jandura}, \citenamefont {Thompson},\ and\ \citenamefont {Pupillo}}]{sven2}%
  \BibitemOpen
  \bibfield  {author} {\bibinfo {author} {\bibfnamefont {S.}~\bibnamefont
  {Jandura}}, \bibinfo {author} {\bibfnamefont {J.~D.}\ \bibnamefont
  {Thompson}},\ and\ \bibinfo {author} {\bibfnamefont {G.}~\bibnamefont
  {Pupillo}},\ }\href {https://doi.org/10.48550/ARXIV.2210.06879} {\bibinfo
  {title} {{Optimizing Rydberg Gates for Logical Qubit Performance}}} (\bibinfo
  {year} {2022}),\ \Eprint {https://arxiv.org/abs/2210.06879} {arXiv:2210.06879
  [quant-ph]} \BibitemShut {NoStop}%
\bibitem [{\citenamefont {Fromonteil}\ \emph {et~al.}(2022)\citenamefont
  {Fromonteil}, \citenamefont {Bluvstein},\ and\ \citenamefont
  {Pichler}}]{fromonteil2022protocols}%
  \BibitemOpen
  \bibfield  {author} {\bibinfo {author} {\bibfnamefont {C.}~\bibnamefont
  {Fromonteil}}, \bibinfo {author} {\bibfnamefont {D.}~\bibnamefont
  {Bluvstein}},\ and\ \bibinfo {author} {\bibfnamefont {H.}~\bibnamefont
  {Pichler}},\ }\href@noop {} {\bibinfo {title} {{Protocols for Rydberg
  entangling gates featuring robustness against quasi-static errors}}}
  (\bibinfo {year} {2022}),\ \Eprint {https://arxiv.org/abs/2210.08824}
  {arXiv:2210.08824 [quant-ph]} \BibitemShut {NoStop}%
\bibitem [{\citenamefont {Johansson}\ \emph {et~al.}(2013)\citenamefont
  {Johansson}, \citenamefont {Nation},\ and\ \citenamefont
  {Nori}}]{JOHANSSON20131234}%
  \BibitemOpen
  \bibfield  {author} {\bibinfo {author} {\bibfnamefont {J.}~\bibnamefont
  {Johansson}}, \bibinfo {author} {\bibfnamefont {P.}~\bibnamefont {Nation}},\
  and\ \bibinfo {author} {\bibfnamefont {F.}~\bibnamefont {Nori}},\ }\bibfield
  {title} {\bibinfo {title} {{QuTiP 2: A Python framework for the dynamics of
  open quantum systems}},\ }\href
  {https://doi.org/https://doi.org/10.1016/j.cpc.2012.11.019} {\bibfield
  {journal} {\bibinfo  {journal} {Comput. Phys. Commun.}\ }\textbf {\bibinfo
  {volume} {184}},\ \bibinfo {pages} {1234} (\bibinfo {year}
  {2013})}\BibitemShut {NoStop}%
\bibitem [{\citenamefont {Seaton}(2002{\natexlab{a}})}]{SEATON2002250}%
  \BibitemOpen
  \bibfield  {author} {\bibinfo {author} {\bibfnamefont {M.}~\bibnamefont
  {Seaton}},\ }\bibfield  {title} {\bibinfo {title} {{FGH}, a code for the
  calculation of {C}oulomb radial wave functions from series expansions},\
  }\href {https://doi.org/https://doi.org/10.1016/S0010-4655(02)00276-X}
  {\bibfield  {journal} {\bibinfo  {journal} {Comput. Phys. Commun.}\ }\textbf
  {\bibinfo {volume} {146}},\ \bibinfo {pages} {250} (\bibinfo {year}
  {2002}{\natexlab{a}})}\BibitemShut {NoStop}%
\bibitem [{\citenamefont {Seaton}(2002{\natexlab{b}})}]{SEATON2002254}%
  \BibitemOpen
  \bibfield  {author} {\bibinfo {author} {\bibfnamefont {M.}~\bibnamefont
  {Seaton}},\ }\bibfield  {title} {\bibinfo {title} {{NUMER}, a code for
  {N}umerov integrations of {C}oulomb functions},\ }\href
  {https://doi.org/https://doi.org/10.1016/S0010-4655(02)00277-1} {\bibfield
  {journal} {\bibinfo  {journal} {Comput. Phys. Commun.}\ }\textbf {\bibinfo
  {volume} {146}},\ \bibinfo {pages} {254} (\bibinfo {year}
  {2002}{\natexlab{b}})}\BibitemShut {NoStop}%
\bibitem [{\citenamefont {Farley}\ and\ \citenamefont
  {Wing}(1981)}]{PhysRevA.23.2397}%
  \BibitemOpen
  \bibfield  {author} {\bibinfo {author} {\bibfnamefont {J.~W.}\ \bibnamefont
  {Farley}}\ and\ \bibinfo {author} {\bibfnamefont {W.~H.}\ \bibnamefont
  {Wing}},\ }\bibfield  {title} {\bibinfo {title} {{Accurate calculation of
  dynamic Stark shifts and depopulation rates of Rydberg energy levels induced
  by blackbody radiation. Hydrogen, helium, and alkali-metal atoms}},\ }\href
  {https://doi.org/10.1103/PhysRevA.23.2397} {\bibfield  {journal} {\bibinfo
  {journal} {Phys. Rev. A}\ }\textbf {\bibinfo {volume} {23}},\ \bibinfo
  {pages} {2397} (\bibinfo {year} {1981})}\BibitemShut {NoStop}%
\bibitem [{\citenamefont {Aymar}\ \emph {et~al.}(1987)\citenamefont {Aymar},
  \citenamefont {Luc-Koenig},\ and\ \citenamefont {Watanabe}}]{Aymar_1987}%
  \BibitemOpen
  \bibfield  {author} {\bibinfo {author} {\bibfnamefont {M.}~\bibnamefont
  {Aymar}}, \bibinfo {author} {\bibfnamefont {E.}~\bibnamefont {Luc-Koenig}},\
  and\ \bibinfo {author} {\bibfnamefont {S.}~\bibnamefont {Watanabe}},\
  }\bibfield  {title} {\bibinfo {title} {{R-matrix calculation of eigenchannel
  multichannel quantum defect parameters for Strontium}},\ }\href
  {https://doi.org/10.1088/0022-3700/20/17/014} {\bibfield  {journal} {\bibinfo
   {journal} {J. Phys. B}\ }\textbf {\bibinfo {volume} {20}},\ \bibinfo {pages}
  {4325} (\bibinfo {year} {1987})}\BibitemShut {NoStop}%
\bibitem [{\citenamefont {Beigang}\ \emph {et~al.}(1982)\citenamefont
  {Beigang}, \citenamefont {Lücke}, \citenamefont {Schmidt}, \citenamefont
  {Timmermann},\ and\ \citenamefont {West}}]{Beigang_1982}%
  \BibitemOpen
  \bibfield  {author} {\bibinfo {author} {\bibfnamefont {R.}~\bibnamefont
  {Beigang}}, \bibinfo {author} {\bibfnamefont {K.}~\bibnamefont {Lücke}},
  \bibinfo {author} {\bibfnamefont {D.}~\bibnamefont {Schmidt}}, \bibinfo
  {author} {\bibfnamefont {A.}~\bibnamefont {Timmermann}},\ and\ \bibinfo
  {author} {\bibfnamefont {P.~J.}\ \bibnamefont {West}},\ }\bibfield  {title}
  {\bibinfo {title} {{One-Photon Laser Spectroscopy of Rydberg Series from
  Metastable Levels in Calcium and Strontium}},\ }\href
  {https://doi.org/10.1088/0031-8949/26/3/007} {\bibfield  {journal} {\bibinfo
  {journal} {Phys. Scr.}\ }\textbf {\bibinfo {volume} {26}},\ \bibinfo {pages}
  {183} (\bibinfo {year} {1982})}\BibitemShut {NoStop}%
\bibitem [{\citenamefont {Wilson}\ \emph {et~al.}(2022)\citenamefont {Wilson},
  \citenamefont {Saskin}, \citenamefont {Meng}, \citenamefont {Ma},
  \citenamefont {Dilip}, \citenamefont {Burgers},\ and\ \citenamefont
  {Thompson}}]{PhysRevLett.128.033201}%
  \BibitemOpen
  \bibfield  {author} {\bibinfo {author} {\bibfnamefont {J.~T.}\ \bibnamefont
  {Wilson}}, \bibinfo {author} {\bibfnamefont {S.}~\bibnamefont {Saskin}},
  \bibinfo {author} {\bibfnamefont {Y.}~\bibnamefont {Meng}}, \bibinfo {author}
  {\bibfnamefont {S.}~\bibnamefont {Ma}}, \bibinfo {author} {\bibfnamefont
  {R.}~\bibnamefont {Dilip}}, \bibinfo {author} {\bibfnamefont {A.~P.}\
  \bibnamefont {Burgers}},\ and\ \bibinfo {author} {\bibfnamefont {J.~D.}\
  \bibnamefont {Thompson}},\ }\bibfield  {title} {\bibinfo {title} {{Trapping
  Alkaline Earth Rydberg Atoms Optical Tweezer Arrays}},\ }\href
  {https://doi.org/10.1103/PhysRevLett.128.033201} {\bibfield  {journal}
  {\bibinfo  {journal} {Phys. Rev. Lett.}\ }\textbf {\bibinfo {volume} {128}},\
  \bibinfo {pages} {033201} (\bibinfo {year} {2022})}\BibitemShut {NoStop}%
\bibitem [{\citenamefont {Blinder}(1984)}]{propagator}%
  \BibitemOpen
  \bibfield  {author} {\bibinfo {author} {\bibfnamefont {S.~M.}\ \bibnamefont
  {Blinder}},\ }\bibfield  {title} {\bibinfo {title} {{Propagators from
  integral representations of Green’s functions for the N‐dimensional
  free‐particle, harmonic oscillator and Coulomb problems}},\ }\href
  {https://doi.org/10.1063/1.526245} {\bibfield  {journal} {\bibinfo  {journal}
  {J. Math. Phys.}\ }\textbf {\bibinfo {volume} {25}},\ \bibinfo {pages} {905}
  (\bibinfo {year} {1984})}\BibitemShut {NoStop}%
\bibitem [{\citenamefont {Bluvstein}\ \emph {et~al.}(2022)\citenamefont
  {Bluvstein} \emph {et~al.}}]{Bluvstein2022}%
  \BibitemOpen
  \bibfield  {author} {\bibinfo {author} {\bibfnamefont {D.}~\bibnamefont
  {Bluvstein}} \emph {et~al.},\ }\bibfield  {title} {\bibinfo {title} {{A
  quantum processor based on coherent transport of entangled atom arrays}},\
  }\href {https://doi.org/10.1038/s41586-022-04592-6} {\bibfield  {journal}
  {\bibinfo  {journal} {Nature (London)}\ }\textbf {\bibinfo {volume} {604}},\
  \bibinfo {pages} {451} (\bibinfo {year} {2022})}\BibitemShut {NoStop}%
\bibitem [{\citenamefont {Khazali}(2023)}]{khazali2023subnanometer}%
  \BibitemOpen
  \bibfield  {author} {\bibinfo {author} {\bibfnamefont {M.}~\bibnamefont
  {Khazali}},\ }\href@noop {} {\bibinfo {title} {{Subnanometer confinement and
  bundling of atoms in a Rydberg empowered optical lattice}}} (\bibinfo {year}
  {2023}),\ \Eprint {https://arxiv.org/abs/2301.04450} {arXiv:2301.04450
  [quant-ph]} \BibitemShut {NoStop}%
\bibitem [{\citenamefont {Steck}(2003)}]{steck}%
  \BibitemOpen
  \bibfield  {author} {\bibinfo {author} {\bibfnamefont {D.~A.}\ \bibnamefont
  {Steck}},\ }\bibfield  {title} {\bibinfo {title} {{Rubidium 87 D Line
  Data}},\ }\href {https://steck.us/alkalidata/rubidium87numbers.1.6.pdf} {\
  (\bibinfo {year} {2003})}\BibitemShut {NoStop}%
\bibitem [{\citenamefont {Couturier}\ \emph {et~al.}(2019)\citenamefont
  {Couturier}, \citenamefont {Nosske}, \citenamefont {Hu}, \citenamefont {Tan},
  \citenamefont {Qiao}, \citenamefont {Jiang}, \citenamefont {Chen},\ and\
  \citenamefont {Weidem{\"u}ller}}]{decayrates1}%
  \BibitemOpen
  \bibfield  {author} {\bibinfo {author} {\bibfnamefont {L.}~\bibnamefont
  {Couturier}}, \bibinfo {author} {\bibfnamefont {I.}~\bibnamefont {Nosske}},
  \bibinfo {author} {\bibfnamefont {F.}~\bibnamefont {Hu}}, \bibinfo {author}
  {\bibfnamefont {C.}~\bibnamefont {Tan}}, \bibinfo {author} {\bibfnamefont
  {C.}~\bibnamefont {Qiao}}, \bibinfo {author} {\bibfnamefont {Y.~H.}\
  \bibnamefont {Jiang}}, \bibinfo {author} {\bibfnamefont {P.}~\bibnamefont
  {Chen}},\ and\ \bibinfo {author} {\bibfnamefont {M.}~\bibnamefont
  {Weidem{\"u}ller}},\ }\bibfield  {title} {\bibinfo {title} {{Measurement of
  the strontium triplet Rydberg series by depletion spectroscopy of ultracold
  atoms}},\ }\href {https://doi.org/10.1103/PhysRevA.99.022503} {\bibfield
  {journal} {\bibinfo  {journal} {Phys. Rev. A}\ }\textbf {\bibinfo {volume}
  {99}},\ \bibinfo {pages} {022503} (\bibinfo {year} {2019})}\BibitemShut
  {NoStop}%
\bibitem [{\citenamefont {Safronova}\ \emph {et~al.}(2013)\citenamefont
  {Safronova}, \citenamefont {Porsev}, \citenamefont {Safronova}, \citenamefont
  {Kozlov},\ and\ \citenamefont {Clark}}]{decayrates2}%
  \BibitemOpen
  \bibfield  {author} {\bibinfo {author} {\bibfnamefont {M.~S.}\ \bibnamefont
  {Safronova}}, \bibinfo {author} {\bibfnamefont {S.~G.}\ \bibnamefont
  {Porsev}}, \bibinfo {author} {\bibfnamefont {U.~I.}\ \bibnamefont
  {Safronova}}, \bibinfo {author} {\bibfnamefont {M.~G.}\ \bibnamefont
  {Kozlov}},\ and\ \bibinfo {author} {\bibfnamefont {C.~W.}\ \bibnamefont
  {Clark}},\ }\bibfield  {title} {\bibinfo {title} {{Blackbody-radiation shift
  in the Sr optical atomic clock}},\ }\href
  {https://doi.org/10.1103/PhysRevA.87.012509} {\bibfield  {journal} {\bibinfo
  {journal} {Phys. Rev. A}\ }\textbf {\bibinfo {volume} {87}},\ \bibinfo
  {pages} {012509} (\bibinfo {year} {2013})}\BibitemShut {NoStop}%
\bibitem [{\citenamefont {Tan}\ \emph {et~al.}(2022)\citenamefont {Tan},
  \citenamefont {Hu}, \citenamefont {Niu}, \citenamefont {Jiang}, \citenamefont
  {Weidem{\"u}ller},\ and\ \citenamefont {Zhu}}]{decayrates3}%
  \BibitemOpen
  \bibfield  {author} {\bibinfo {author} {\bibfnamefont {C.}~\bibnamefont
  {Tan}}, \bibinfo {author} {\bibfnamefont {F.}~\bibnamefont {Hu}}, \bibinfo
  {author} {\bibfnamefont {Z.}~\bibnamefont {Niu}}, \bibinfo {author}
  {\bibfnamefont {Y.}~\bibnamefont {Jiang}}, \bibinfo {author} {\bibfnamefont
  {M.}~\bibnamefont {Weidem{\"u}ller}},\ and\ \bibinfo {author} {\bibfnamefont
  {B.}~\bibnamefont {Zhu}},\ }\bibfield  {title} {\bibinfo {title}
  {{Measurements of Dipole Moments for the
  $\mathrm{5s5p^3P_1–5s}n\mathrm{s^3S_1}$ Transitions via Autler-Townes
  Spectroscopy}},\ }\href {https://doi.org/10.1088/0256-307X/39/9/093202}
  {\bibfield  {journal} {\bibinfo  {journal} {Chin. Phys. Lett.}\ }\textbf
  {\bibinfo {volume} {39}},\ \bibinfo {pages} {093202} (\bibinfo {year}
  {2022})}\BibitemShut {NoStop}%
\bibitem [{\citenamefont {Zhang}\ \emph {et~al.}(2012)\citenamefont {Zhang},
  \citenamefont {Gill}, \citenamefont {Isenhower}, \citenamefont {Walker},\
  and\ \citenamefont {Saffman}}]{PhysRevA.85.042310}%
  \BibitemOpen
  \bibfield  {author} {\bibinfo {author} {\bibfnamefont {X.~L.}\ \bibnamefont
  {Zhang}}, \bibinfo {author} {\bibfnamefont {A.~T.}\ \bibnamefont {Gill}},
  \bibinfo {author} {\bibfnamefont {L.}~\bibnamefont {Isenhower}}, \bibinfo
  {author} {\bibfnamefont {T.~G.}\ \bibnamefont {Walker}},\ and\ \bibinfo
  {author} {\bibfnamefont {M.}~\bibnamefont {Saffman}},\ }\bibfield  {title}
  {\bibinfo {title} {{Fidelity of a Rydberg-blockade quantum gate from
  simulated quantum process tomography}},\ }\href
  {https://doi.org/10.1103/PhysRevA.85.042310} {\bibfield  {journal} {\bibinfo
  {journal} {Phys. Rev. A}\ }\textbf {\bibinfo {volume} {85}},\ \bibinfo
  {pages} {042310} (\bibinfo {year} {2012})}\BibitemShut {NoStop}%
\bibitem [{\citenamefont {Booth}\ \emph {et~al.}(2018)\citenamefont {Booth},
  \citenamefont {Isaacs},\ and\ \citenamefont {Saffman}}]{PhysRevA.97.012515}%
  \BibitemOpen
  \bibfield  {author} {\bibinfo {author} {\bibfnamefont {D.~W.}\ \bibnamefont
  {Booth}}, \bibinfo {author} {\bibfnamefont {J.}~\bibnamefont {Isaacs}},\ and\
  \bibinfo {author} {\bibfnamefont {M.}~\bibnamefont {Saffman}},\ }\bibfield
  {title} {\bibinfo {title} {{Reducing the sensitivity of Rydberg atoms to dc
  electric fields using two-frequency ac field dressing}},\ }\href
  {https://doi.org/10.1103/PhysRevA.97.012515} {\bibfield  {journal} {\bibinfo
  {journal} {Phys. Rev. A}\ }\textbf {\bibinfo {volume} {97}},\ \bibinfo
  {pages} {012515} (\bibinfo {year} {2018})}\BibitemShut {NoStop}%
\bibitem [{\citenamefont {Sevinçli}\ and\ \citenamefont
  {Pohl}(2014)}]{sevincli_2014}%
  \BibitemOpen
  \bibfield  {author} {\bibinfo {author} {\bibfnamefont {S.}~\bibnamefont
  {Sevinçli}}\ and\ \bibinfo {author} {\bibfnamefont {T.}~\bibnamefont
  {Pohl}},\ }\bibfield  {title} {\bibinfo {title} {{Microwave control of
  Rydberg atom interactions}},\ }\href
  {https://doi.org/10.1088/1367-2630/16/12/123036} {\bibfield  {journal}
  {\bibinfo  {journal} {New J. Phys.}\ }\textbf {\bibinfo {volume} {16}},\
  \bibinfo {pages} {123036} (\bibinfo {year} {2014})}\BibitemShut {NoStop}%
\bibitem [{\citenamefont {Weber}\ \emph {et~al.}(2017)\citenamefont {Weber},
  \citenamefont {Tresp}, \citenamefont {Menke}, \citenamefont {Urvoy},
  \citenamefont {Firstenberg}, \citenamefont {B{\"u}chler},\ and\ \citenamefont
  {Hofferberth}}]{Weber2017}%
  \BibitemOpen
  \bibfield  {author} {\bibinfo {author} {\bibfnamefont {S.}~\bibnamefont
  {Weber}}, \bibinfo {author} {\bibfnamefont {C.}~\bibnamefont {Tresp}},
  \bibinfo {author} {\bibfnamefont {H.}~\bibnamefont {Menke}}, \bibinfo
  {author} {\bibfnamefont {A.}~\bibnamefont {Urvoy}}, \bibinfo {author}
  {\bibfnamefont {O.}~\bibnamefont {Firstenberg}}, \bibinfo {author}
  {\bibfnamefont {H.~P.}\ \bibnamefont {B{\"u}chler}},\ and\ \bibinfo {author}
  {\bibfnamefont {S.}~\bibnamefont {Hofferberth}},\ }\bibfield  {title}
  {\bibinfo {title} {{Tutorial: Calculation of Rydberg interaction
  potentials}},\ }\href {https://doi.org/10.1088/1361-6455/aa743a} {\bibfield
  {journal} {\bibinfo  {journal} {J. Phys. B}\ }\textbf {\bibinfo {volume}
  {50}},\ \bibinfo {pages} {133001} (\bibinfo {year} {2017})}\BibitemShut
  {NoStop}%
\bibitem [{\citenamefont {Yang}\ \emph {et~al.}(2022)\citenamefont {Yang},
  \citenamefont {Nie}, \citenamefont {Xin}, \citenamefont {Lu},\ and\
  \citenamefont {Li}}]{yang2022quantum}%
  \BibitemOpen
  \bibfield  {author} {\bibinfo {author} {\bibfnamefont {X.}~\bibnamefont
  {Yang}}, \bibinfo {author} {\bibfnamefont {X.}~\bibnamefont {Nie}}, \bibinfo
  {author} {\bibfnamefont {T.}~\bibnamefont {Xin}}, \bibinfo {author}
  {\bibfnamefont {D.}~\bibnamefont {Lu}},\ and\ \bibinfo {author}
  {\bibfnamefont {J.}~\bibnamefont {Li}},\ }\href@noop {} {\bibinfo {title}
  {{Quantum Control for Time-dependent Noise by Inverse Geometric
  Optimization}}} (\bibinfo {year} {2022}),\ \Eprint
  {https://arxiv.org/abs/2205.02515} {arXiv:2205.02515 [quant-ph]} \BibitemShut
  {NoStop}%
\bibitem [{\citenamefont {Colmenar}\ and\ \citenamefont
  {Kestner}(2022)}]{PhysRevA.106.032611}%
  \BibitemOpen
  \bibfield  {author} {\bibinfo {author} {\bibfnamefont {R.~K.~L.}\
  \bibnamefont {Colmenar}}\ and\ \bibinfo {author} {\bibfnamefont {J.~P.}\
  \bibnamefont {Kestner}},\ }\bibfield  {title} {\bibinfo {title} {{Reverse
  engineering of one-qubit filter functions with dynamical invariants}},\
  }\href {https://doi.org/10.1103/PhysRevA.106.032611} {\bibfield  {journal}
  {\bibinfo  {journal} {Phys. Rev. A}\ }\textbf {\bibinfo {volume} {106}},\
  \bibinfo {pages} {032611} (\bibinfo {year} {2022})}\BibitemShut {NoStop}%
\bibitem [{\citenamefont {Khazali}\ and\ \citenamefont
  {Lechner}(2023)}]{khazali2023electron}%
  \BibitemOpen
  \bibfield  {author} {\bibinfo {author} {\bibfnamefont {M.}~\bibnamefont
  {Khazali}}\ and\ \bibinfo {author} {\bibfnamefont {W.}~\bibnamefont
  {Lechner}},\ }\href@noop {} {\bibinfo {title} {{Electron cloud design for
  Rydberg multi-qubit gates}}} (\bibinfo {year} {2023}),\ \Eprint
  {https://arxiv.org/abs/2111.01581} {arXiv:2111.01581 [quant-ph]} \BibitemShut
  {NoStop}%
\end{thebibliography}%

\end{document}